\documentclass{aa}

\usepackage{txfonts}
\usepackage{tabularx}
\usepackage{subfigure}  
\usepackage{graphicx,amsmath}
\usepackage{amssymb}
\usepackage{color}
\usepackage{natbib}		
\usepackage{url}
\usepackage{multirow}
\usepackage[para,online,flushleft]{threeparttable}   
\usepackage{soul,mathabx}
\bibpunct{(}{)}{;}{a}{}{,} 
\usepackage{lscape}
\usepackage{adjustbox,textcomp}
\usepackage{dblfloatfix}

\usepackage{hyperref}   
\usepackage{braket}     
\usepackage{upgreek}    
\usepackage{array,multirow,booktabs}

\newcommand{\appropto}{\mathrel{\vcenter{
  \offinterlineskip\halign{\hfil$##$\cr
    \propto\cr\noalign{\kern2pt}\sim\cr\noalign{\kern-2pt}}}}}

\begin{document}

   \title{DREAM II. The spin--orbit angle distribution of close-in exoplanets under the lens of tides}

   \author{O.~Attia
      \and V.~Bourrier
      \and J.-B.~Delisle
      \and P.~Eggenberger}

   \institute{Observatoire Astronomique de l'Universit\'e de Gen\`eve, Chemin Pegasi 51b, CH-1290 Versoix, Switzerland\\
              \email{omar.attia@unige.ch}
             }

\authorrunning{O.~Attia et al.}
\titlerunning{DREAM II}

\date{Received XXX; accepted XXX}



\abstract
   {
   The spin--orbit angle, or obliquity, is a powerful observational marker that allows us to access the dynamical history of exoplanetary systems. For this study, we have examined the distribution of spin--orbit angles for close-in exoplanets and put it in a statistical context of tidal interactions between planets and their host stars. We confirm the previously observed trends between the obliquity and physical quantities directly connected to tides, namely the stellar effective temperature, the planet-to-star mass ratio, and the scaled orbital distance. We further devised a tidal efficiency factor $\tau$ combining critical parameters that control the strength of tidal effects and used it to corroborate the strong link between the spin--orbit angle distribution and tidal interactions. In particular, we developed a readily usable formula $\theta \left( \tau \right)$ to estimate the probability that a system is misaligned, which will prove useful in global population studies. By building a robust statistical framework, we reconstructed the distribution of the three-dimensional spin--orbit angles, allowing for a sample of nearly 200 true obliquities to be analyzed for the first time. This realistic distribution maintains the sky-projected trends, and additionally hints toward a striking pileup of truly aligned systems. In fact, we show that the fraction of aligned orbits could be underestimated in classical analyses of sky-projected obliquities due to an observational bias toward misaligned systems. The comparison between the full population and a pristine subsample unaffected by tidal interactions suggests that perpendicular architectures are resilient toward tidal realignment, providing evidence that orbital misalignments are sculpted by disruptive dynamical processes that preferentially lead to polar orbits. On the other hand, star--planet interactions seem to efficiently realign or quench the formation of any tilted configuration other than for polar orbits, and in particular for antialigned orbits. Observational and theoretical efforts focused on these pristine systems are encouraged in order to study primordial mechanisms shaping orbital architectures, which are unaltered by tidal effects. \\
   }


\maketitle


\section{Introduction}
\label{sec:intro}

The detection of the first exoplanet around a main sequence star \citep{Mayor1995} sparked a revolution in planetary science. Instead of orbiting far from its star similar to gas giants in our Solar System, the discovered ``hot Jupiter'' was found to orbit ten times closer than Mercury around the Sun, challenging our understanding of planetary formation and evolution. Nearly 30 years later, a wealth of exoplanets evolving on very short orbits have been discovered, enabling comprehensive studies on this still intriguing class of distant worlds. As a matter of fact, the population of close-in exoplanets shows a wider range of orbital characteristics that had been expected from our initial interpretations of the Solar System \citep[e.g.,][]{Winn2015}. 

\begin{figure*}
\includegraphics[width=0.9\textwidth]{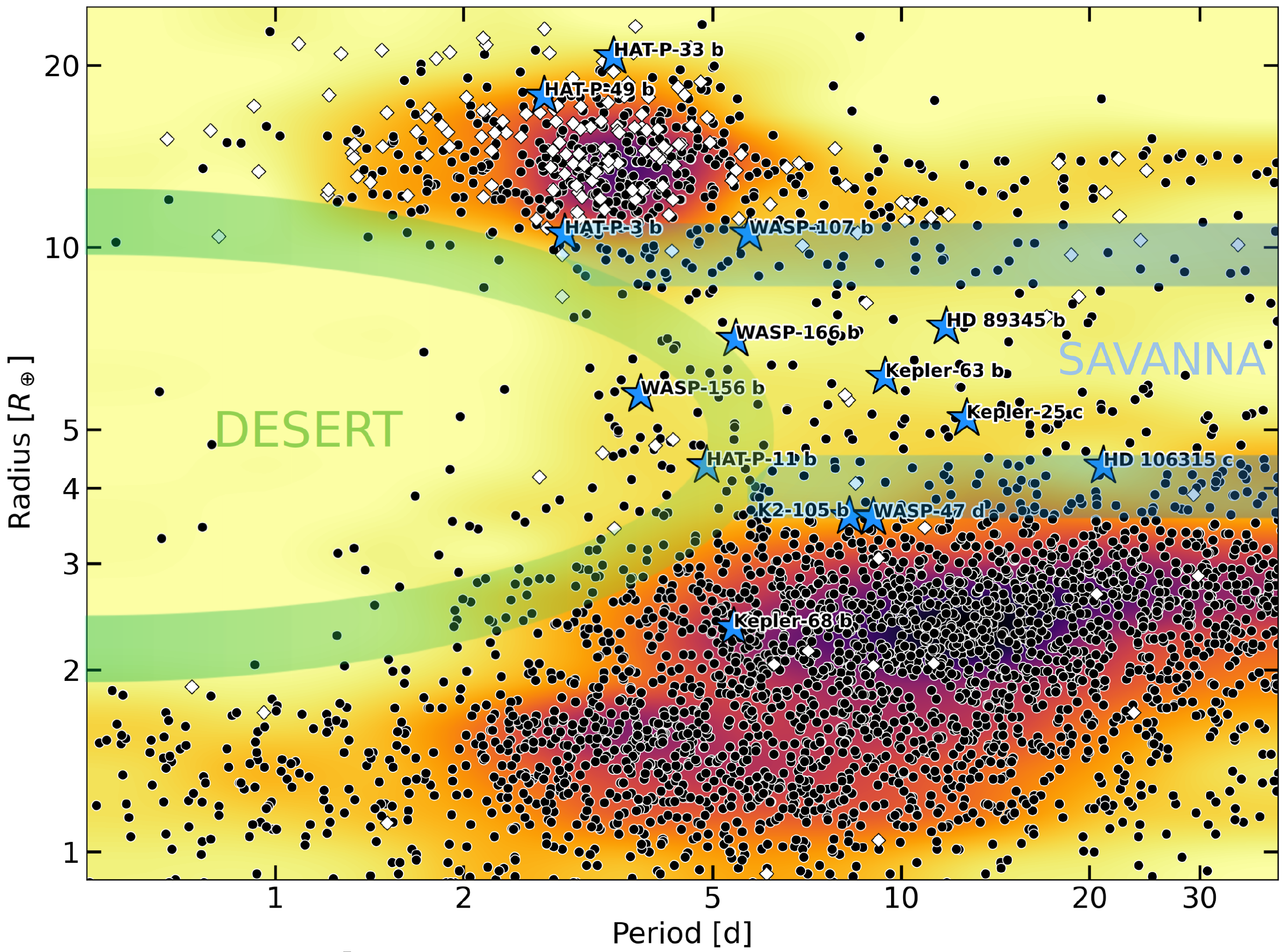}
\centering
\caption{Distribution of close-in exoplanets as a function of their radius and orbital period, featuring the Neptune desert and savanna. White diamonds indicate exoplanets with measured spin--orbit angles. Blue stars highlight planets in the survey described in DREAM I.}
\label{fig:population}
\end{figure*}

One of the features illustrating this diversity is the ``hot Neptune desert,'' with a striking paucity of intermediate-sized planets with short periods \citep[$P < 3$ d, e.g.,][see also Fig.~\ref{fig:population}]{Lecav2007,Beauge2013,Lundkvist2016}, as well as a ``Neptune savanna,'' identified as a milder deficit at longer periods \citep{Bourrier2023}. It is thought that orbital migration plays an important role in shaping the observed demographics of close-in planets and the desert in particular \citep[e.g.,][]{Mazeh2016,Bourrier_2018_Nat,Attia2021}, although it remains to be explored how the various migration processes shape the orbital distribution of Neptunian worlds across the savanna and desert. Different mechanisms have indeed been proposed to transport planets from their presumed birthplace to their current location. Some of them would preserve the relative orientation between the planetary orbit and the stellar equator \citep[e.g.,][]{Lin1996,Zhou2020,Mann2020}, while others could disrupt this alignment \citep[e.g.,][]{Fabrycky2007,Naoz2011,Nelson2017}, offering us the possibility to distinguish between different theories for the origin of close-in systems.

For this reason, this so-called spin--orbit angle (or obliquity) represents a crucial observational tracer for past dynamical histories. This angle can be accessed when the probed planet transits its host star, creating detectable anomalies in the spectral absorption lines of the eclipsed star originating from the partial occultation of the rotating photosphere. This phenomenon, dubbed the Rossiter--McLaughlin (RM) effect \citep{Rossiter1924,McLaughlin1924}, has been a successful tool over the past decades to unveil a plethora of surprisingly misaligned orbits \citep[see][for a recent review]{Albrecht2022}. Yet, it is a difficult task to trace all the steps that occurred during the past secular evolution of close-in planets, especially since their short final orbits make them vulnerable to strong tidal interactions with their host star, which progressively smooth out signs of disruptive dynamical histories \citep[e.g.,][]{Barker2010}.

With the present work, we contribute to this global effort by conducting a statistical study of the entire sample of known obliquity measurements. This is the second article of the DREAM series, whose objectives are to better understand the origins and evolution of close-in planets, particularly those closely linked to the hot Neptune desert. In DREAM I \citep{Bourrier2023}, we determined the orbital architecture of several planets spanning the borders of the desert and savanna (Fig.~\ref{fig:population}) using the Rossiter--McLaughlin revolutions (RMR) technique \citep{Bourrier2021}. For this study, we have put this sample in a larger statistical framework, looked for possible trends, and identified interesting links with star--planet interactions.

\section{Observational trends: A link with tides}
\label{sec:trends_tides}


\subsection{Constructing the obliquity sample}

The survey described in detail in DREAM I \citep{Bourrier2023} provides a substantial revision to the systems with a characterized spin--orbit angle. The sky-projected obliquity $\lambda$ was successfully measured for 11 targets. Additionally, three host stars had known stellar inclinations, and it could be constrained for four more, allowing us to derive the true obliquity $\psi$ in seven systems. In this section, we put our sample in the context of the close-in planet population, investigate known trends for spin--orbit angles, and search for outliers from these trends that could hint at interesting dynamical histories. To this effect, we collect spin--orbit angle measurements from the TEPCat catalog \citep{Southworth2011} after a careful inspection of the accuracy of all their sources, and combine the NASA Exoplanet Archive\footnote{\url{https://exoplanetarchive.ipac.caltech.edu/}} and Exoplanet.eu\footnote{\url{http://exoplanet.eu/}} for all other relevant parameters. In total, our sample consists in 196 spin--orbit measurements, 42 of which are three-dimensional (3D), the remaining 154 being sky-projected. All of the obliquity measurements used in this work are summarized in App.~\ref{app:sample}.

Being interested in the geometrical configuration of exoplanet systems, and not the spin--orbit angle value per se, we fold the values to [0 ; 180]$^\circ$ (e.g., retrograde orbits are then grouped together irrespective of their orientation). In doing so, we follow the same scheme as the state-of-the-art review of \citet{Albrecht2022}. Because of the possible biases on the obliquity associated with the classical RM method, we choose to use values derived through the RMR analysis for the sample studied in DREAM I, even when values had already been published. In the following analysis, we define misaligned orbits with $\lambda$ or $\psi > 30^\circ$ as a conservative threshold given the typical uncertainties on RM measurements. We plot spin--orbit angle measurements as a function of various parameters to investigate possible trends. On top of these figures, we draw a bar plot counting the percentage of misaligned systems within each $x$-axis bin, and we adapt the bin sizes to have roughly equivalent sample sizes in each of them. In this section, we only use the obliquities of App.~\ref{app:sample} coming from the literature, in order to perform a reference-point analysis. We note again that many spin--orbit angles in this sample only have sky-projected measurements, which could bias the interpretation of their distribution. We address these caveats in Sect.~\ref{sec:estim_psi}.


\subsection{The Kraft break}

\begin{figure*}
\includegraphics[width=0.96\textwidth]{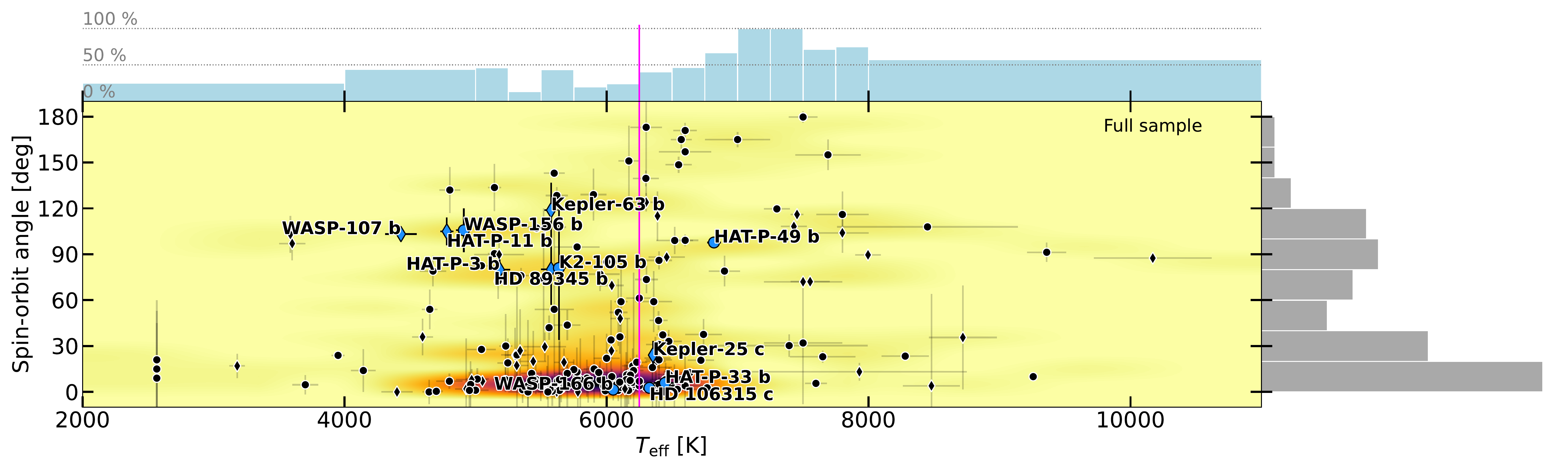}
\includegraphics[width=0.48\textwidth]{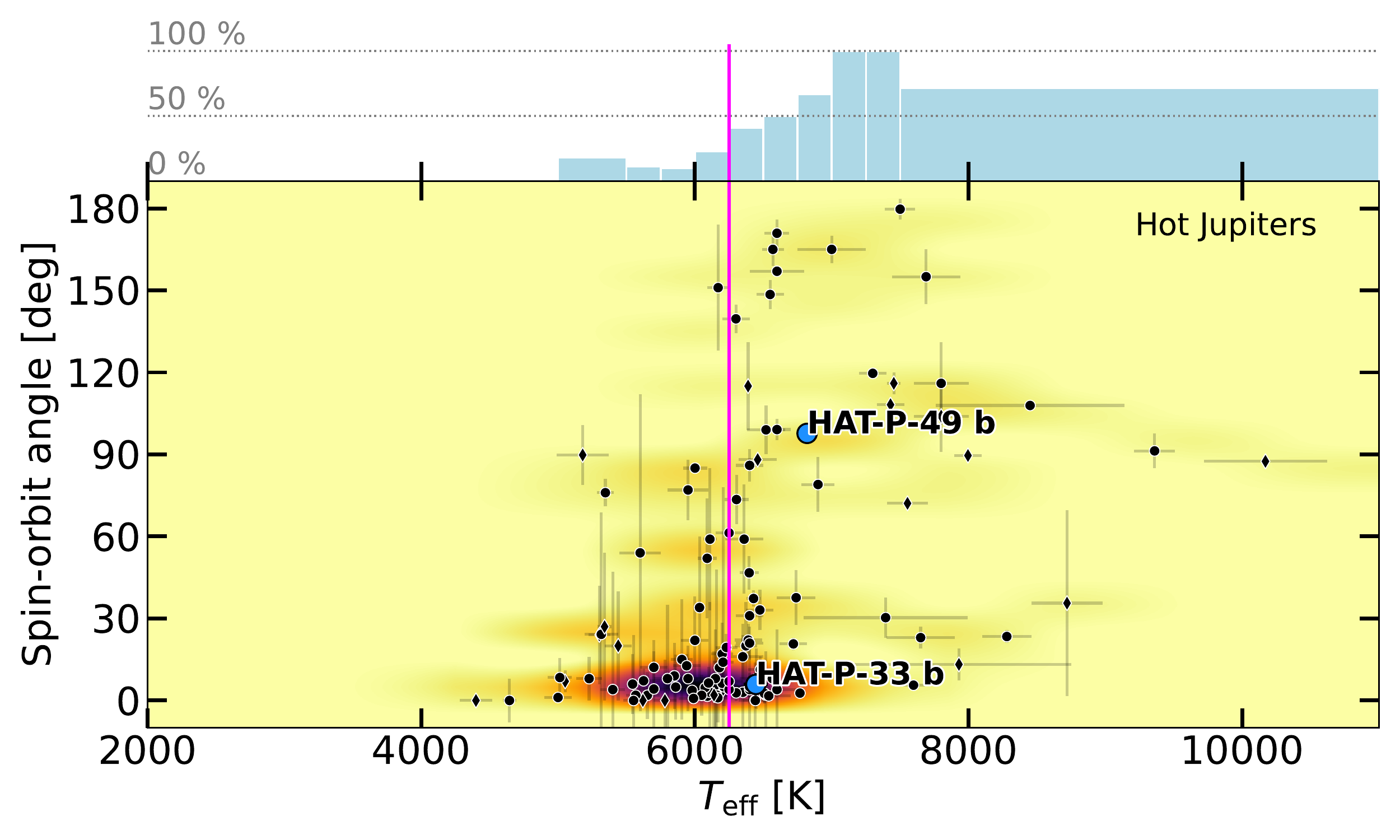}
\includegraphics[width=0.48\textwidth]{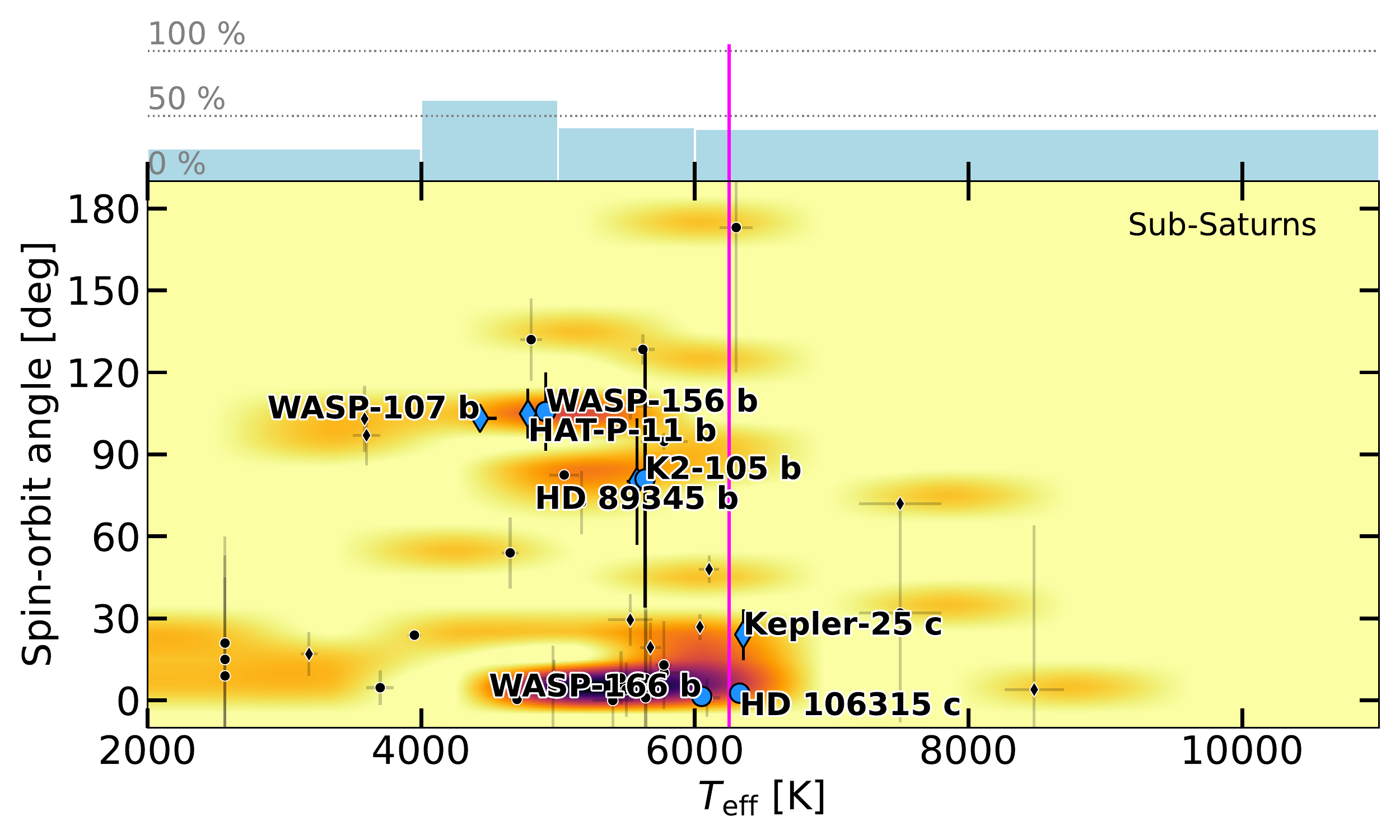}
\centering
\caption{Distribution of spin--orbit angles as a function of the stellar effective temperature. Circles represent projected obliquities $\lambda$ and diamonds true obliquities $\psi$. Exoplanets from the survey described in DREAM I are highlighted as blue symbols. The blue bars on top of the plots count the percentage of misaligned systems ($\lambda$ or $\psi > 30^\circ$) within each $T_{\rm eff}$ bin. The top panel encompasses all systems with a known spin--orbit angle, whereas the bottom panels are subsets with only hot Jupiters (left) and sub-Saturns (right). The gray histogram on the right of the top plot counts the number of measurements within each spin--orbit angle bin. The color map represents a smoothed number-density of planets to guide the eye. The Kraft break ($T_{\rm eff}=6250$ K) is shown as a magenta vertical line.}
\label{fig:Teff_hist}
\end{figure*}

We first look at the possible correlation between the spin--orbit angle and stellar effective temperature ($T_{\rm eff}$, Fig.~\ref{fig:Teff_hist}). A $T_{\rm eff}$ trend was first noted by \citet{winn2010a} for hot Jupiters alone, mainly because this class of planets was the only one accessible to RM measurements. Even though the sample size has nearly quintupled since this seminal study, our analysis confirms this trend in line with \citet{Brown2017}, \citet{Triaud2018}, and \citet{Albrecht2022}, for example. Planets around hot stars \citep[$T_{\rm eff} > 6250$ K, or the Kraft break,][]{Kraft1967} tend to be more misaligned than around cold stars. This is linked to the transition to F8 stars and the disappearance of the stellar convective envelope at roughly that temperature threshold \citep{Kippenhahn2013}. Indeed, cold stars with deep convective envelopes undergo magnetic braking and rotate slower, which hastens the effects of tidal dissipation and allows for a more efficient realignment of the system \citep[e.g.,][]{Hansen2012,Valsecchi2014}.

Compared to \citet{winn2010a}, the larger sample of spin--orbit angles that is now available allows us to draw a more global picture on their distribution. In particular, one can see in Fig.~\ref{fig:Teff_hist} (right of top panel) that it is far from uniform. Aligned systems seem to be dominant, but contrary to what was suggested by \citet{winn2010a}, misaligned systems do not randomly span the full range of obliquities. Instead, they seem to pile up on polar ($\sim 90^\circ$) orbits, a feature that was unveiled by \citet{Albrecht2021} and deemed unlikely to be a statistical fluke. Our analysis agrees with their results and further hints at the existence of a mechanism exciting spin--orbit angles that has yet to be assessed, which preferentially leads to polar orbits.

From the top panel of Fig.~\ref{fig:Teff_hist}, one can also notice that despite the $\lambda - T_{\rm eff}$ trend, many planets around cold stars are on substantially misaligned orbits, primarily polar ones. In particular, this is the case for all the planets of our survey that orbit cold stars, except for WASP-166 b. These apparent outliers can in reality still be explained by tidal interactions. Breaking down the sample into hot Jupiters and sub-Saturns separately (Fig.~\ref{fig:Teff_hist}, lower panels), the misalignment trend is even clearer for the former while the latter shows a flat distribution. Because of their larger mass, hot Jupiters raise more efficient tides, which is why the trend is all the more accentuated for this subsample. Our apparent outliers turn out to be part of the sub-Saturn group, for which no trend with $T_{\rm eff}$ is seen since these lower-mass planets raise less efficient tides even around cold stars. 

An apparent tension might arise when comparing this last result to \citet{Morton2014}, \citet{Mazeh2015}, and \citet{Louden2021}, who suggested that the Kraft break persists even for smaller planets based on a statistical argument regarding stellar inclinations of the Kepler sample. However, our sub-Saturn sample still consists in large, and rather lone planets, thus mainly affected by tides with the host star. On the other hand, their analyses concerns Kepler planets, smaller in size and generally part of compact multiplanet systems, for which gravitational interactions between the different planets is dominant over tides. Hence, our study can be seen as complementary to theirs, looking into exoplanets affected by a different class of mechanisms.


\subsection{Impact of mass and separation}

\begin{figure*}
\includegraphics[width=0.48\textwidth]{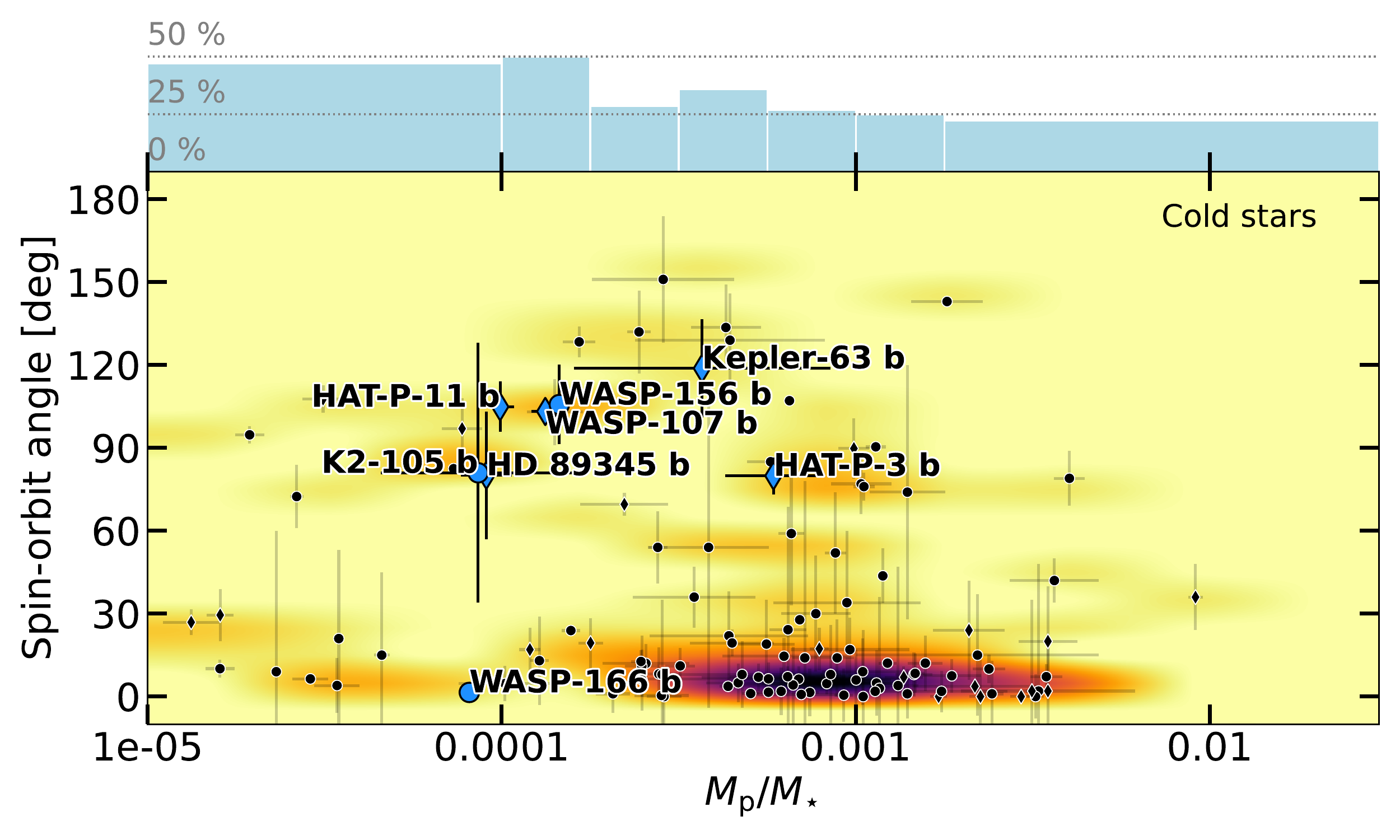}
\includegraphics[width=0.48\textwidth]{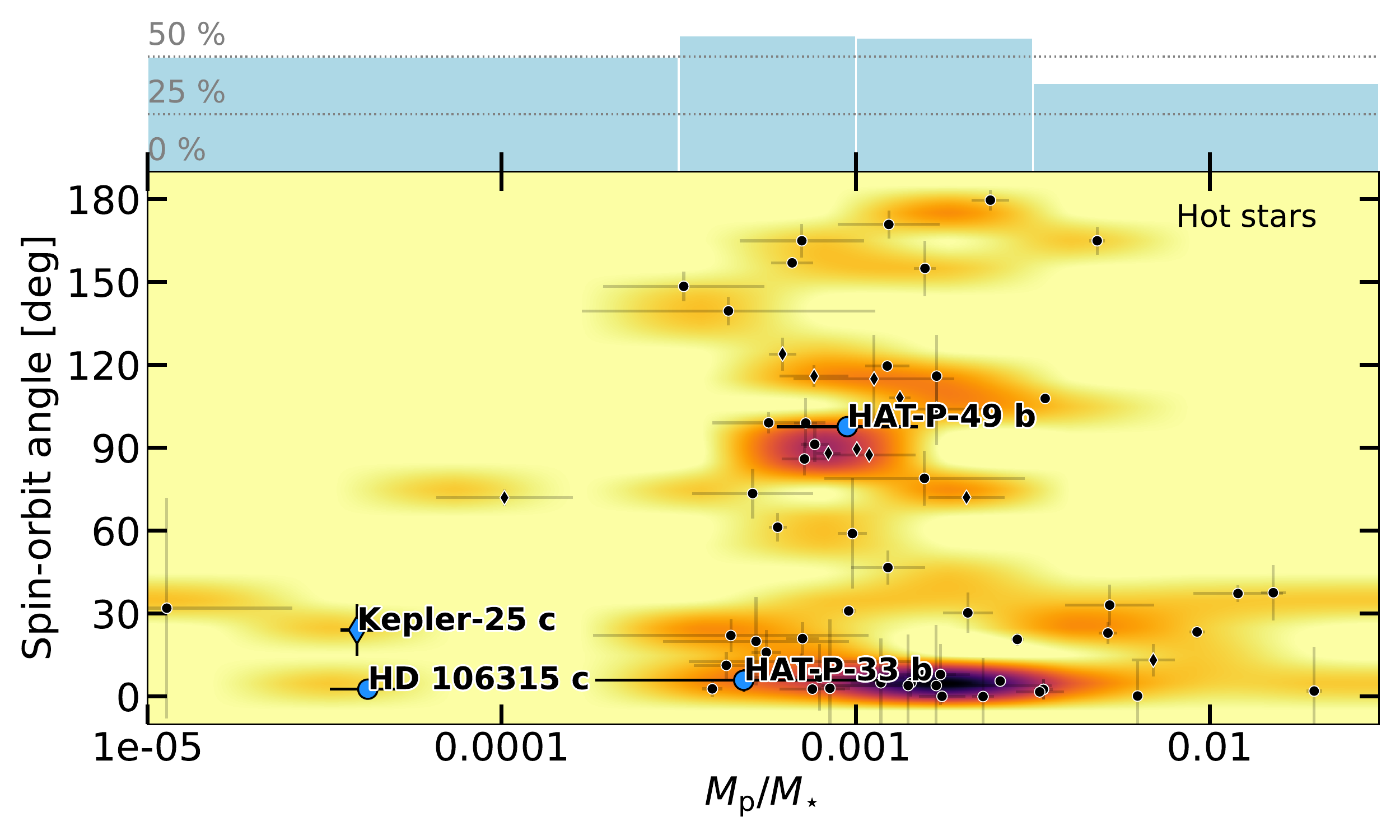}
\centering
\caption{Distribution of spin--orbit angles as a function of the planet-to-star mass ratio, subdivided into systems with a cold ($T_{\rm eff} < 6250$ K, left) or hot ($T_{\rm eff} > 6250$ K, right) host star. Same color, symbol, and histogram schemes as in Fig.~\ref{fig:Teff_hist}.}
\label{fig:MpMs_hist}
\end{figure*}

To further investigate the impact of planetary mass in realigning the stellar spin-axis, we show the spin--orbit angle distribution as a function of the planet-to-star mass ratio ($M_{\rm p}/M_\star$, Fig.~\ref{fig:MpMs_hist}). Scrutinizing the cold star sample in isolation, we see a downward trend of misalignment with $M_{\rm p}/M_\star$. More massive planets tend to be more aligned, a trend already noted by \citet{hebrard2011b} with a smaller sample. This is understood because the higher the mass of the planet relative to the mass of the star, the faster its tides can align the stellar and orbital angular momentum vectors \citep[e.g.,][]{Barker2010,Dawson2014}. In contrast, hot stars, for which tidal realignment is inefficient, expectedly showcase a flat misalignment distribution.

\begin{figure*}
\includegraphics[width=0.48\textwidth]{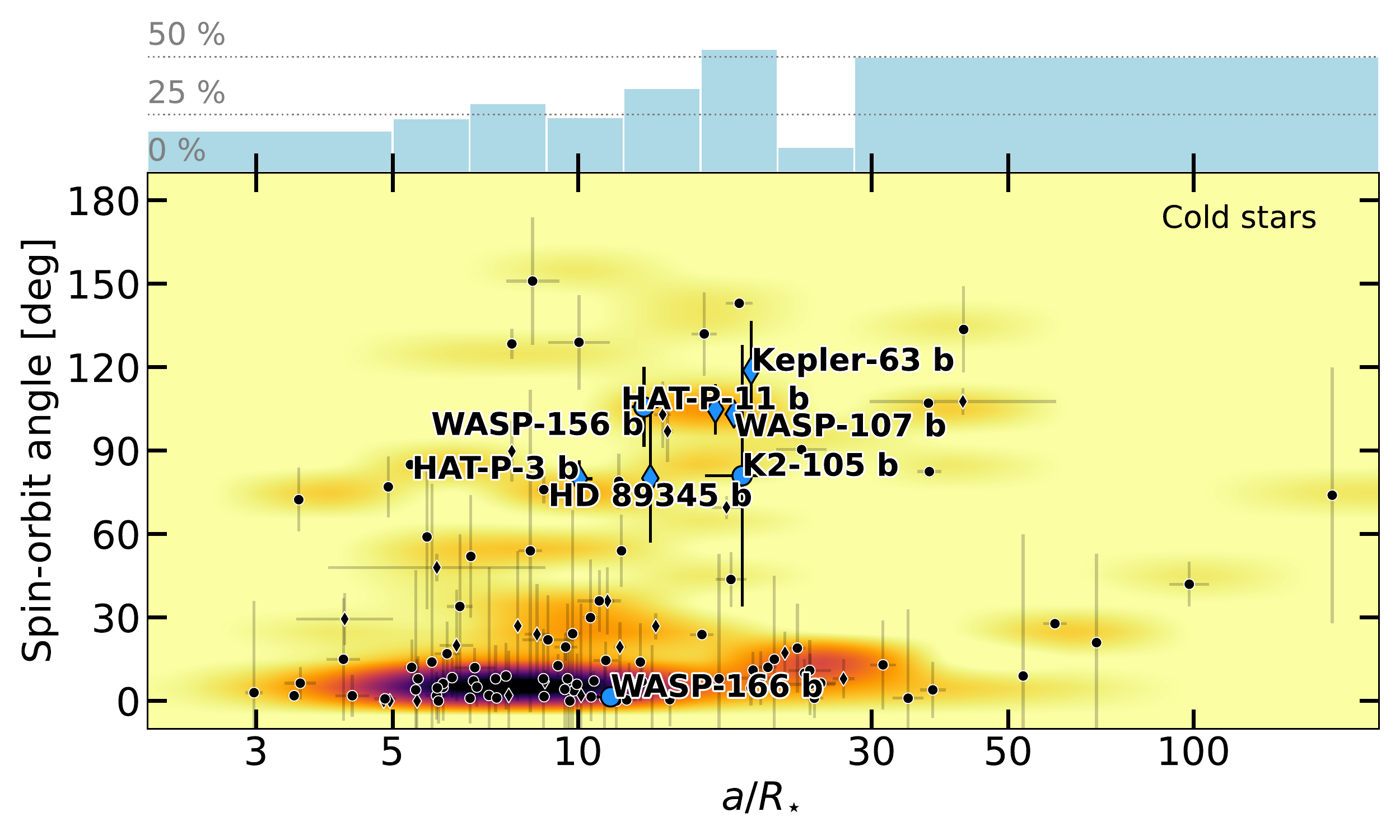}
\includegraphics[width=0.48\textwidth]{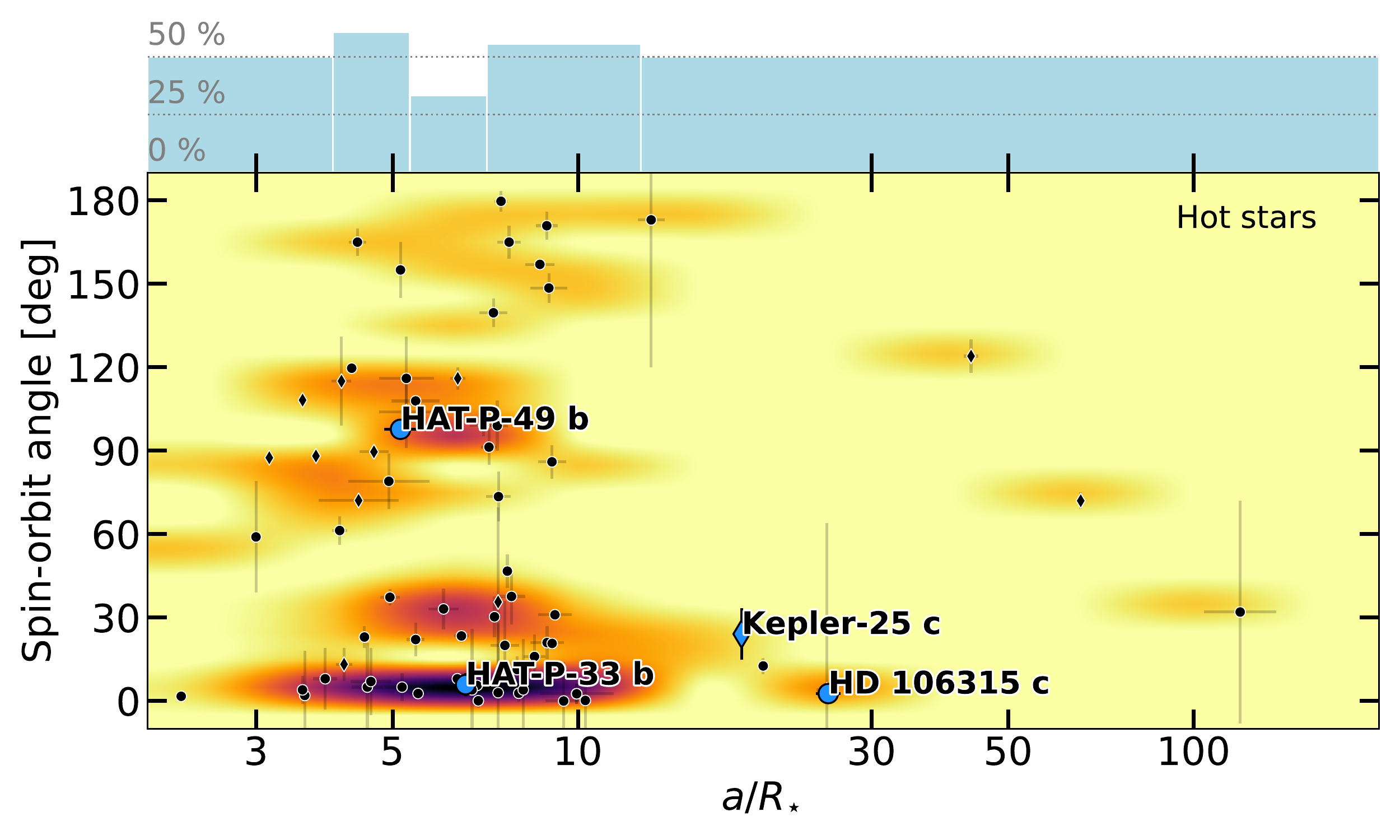}
\centering
\caption{Distribution of spin--orbit angles as a function of the semi-major axis-to-stellar radius ratio, subdivided into systems with a cold ($T_{\rm eff} < 6250$ K, left) or hot ($T_{\rm eff} > 6250$ K, right) host star. Same color, symbol, and histogram schemes as in Fig.~\ref{fig:Teff_hist}.}
\label{fig:aRs_hist}
\end{figure*}

The major impact of tides is further confirmed by the distribution of spin--orbit angles as a function of the scaled separation ($a/R_\star$, Fig.~\ref{fig:aRs_hist}). Again, for cold stars, which efficiently realign their systems, there is an upward misalignment trend with the semi-major axis \citep[as broadly seen in][]{Anderson2015}. Systems with a short separation are more impacted by tides \citep[e.g.,][]{Barker2010,Dawson2014} and get realigned more easily while planets further out do not feel tidal effects and showcase a broader spin--orbit angle distribution. We note the surprisingly aligned bin at $a/R_\star \lesssim 30$ (Fig.~\ref{fig:aRs_hist}, left panel), which we hypothesize could trace the final orbit reached by planets that formed locally or underwent disk-driven migration in a coplanar protoplanetary disk. On the other hand, we expectedly see a flat distribution for hot stars irrespective of the orbital distance.

\section{Quantifying tidal realignment}

The previous investigated trends, although purely qualitative, show promising links between the spin--orbit architecture of close-in planet systems and the intensity of tidal interactions within them, paving the way for a more in-depth analysis. In this section, we set the analytical framework that will allow us to rigorously quantify the connection between the obliquity distribution and tidal theory.


\subsection{Tidal efficiency factor}

Our aim here is to combine the various parameters involved in the strength of tidal effects (Sect.~\ref{sec:trends_tides}) to design a global tidal efficiency factor. We employ the formalism of equilibrium tides, descried in detail in \citet{Zahn1966}, \citet{Alexander1976}, \citet{Zahn1977}, and \citet{Zahn1989}. As seen in \citet{Zahn1977}, for example, the characteristic timescale $t_a$ for a change in the semi-major axis $a$ due to equilibrium tides is proportional to

\begin{equation}
\label{eq:ta}
 \frac{1}{t_a} = \left| \dot{a}/a \right|_\mathrm{eq} \; \propto \; \frac{M_\mathrm{p}}{M_\star}\left(1 + \frac{M_\mathrm{p}}{M_\star}\right) 
 \left( \frac{R_{\star}}{a} \right)^8,
\end{equation}

\noindent The corresponding timescale $t_\psi$ for a change in obliquity is then given by \citep[e.g.,][]{Lai2012}

\begin{equation}
 \frac{1}{t_\psi} \simeq \frac{L_\mathrm{p}}{2L_\star} \frac{1}{t_a},
\end{equation}

\noindent where $L_{\rm p}$ and $L_\star$ are the orbital angular momentum and the stellar spin angular momentum, respectively. The latter quantities are proportional to $L_{\rm p} \; \propto \; M_{\rm p} a^2 / P$ and $L_\star \; \propto \; M_\star R_\star^2 / P_{\rm eq}$ respectively, where $P$ is the orbital period and $P_{\rm eq}$ the stellar rotation period. Consequently, linking this obliquity variation timescale to measurable parameters yields

\begin{equation}
\label{eq:to}
  \frac{1}{t_\psi} \; \propto \; \frac{P_\mathrm{eq}}{P} \left(\frac{M_\mathrm{p}}{M_\star}\right)^2 \left(1 + \frac{M_\mathrm{p}}{M_\star}\right) \left(\frac{a}{R_\star}\right)^{-6}.
\end{equation}

\noindent Our main objective is to quantify tidal dissipation efficiency for the targets in our sample using Eq.~(\ref{eq:to}), with order of magnitude comparison. Thus, the $1 + M_{\rm p}/M_\star$ term can safely be discarded as it stays close to unity for planetary systems. Moreover, the $P_{\rm eq}/P$ term can be neglected because these two periods usually remain within the same order of magnitude, while the other ratios in Eq.~(\ref{eq:to}) strongly vary from one system to another, especially when raised to high powers. To quantify the variability in orders of magnitude of the different factors intervening in Eq.~(\ref{eq:to}), we compute the standard deviation of their logarithms for all systems with precise (relative error $< 10\%$) measurements. We find $\sigma \left( \log (P_{\rm eq}/P) \right) = 0.78$, $\sigma \left( \log (M_{\rm p}/M_\star)^2 \right) = 2.16$, and $\sigma \left( \log (a/R_\star)^{-6} \right) = 4.19$, confirming that $P_{\rm eq}/P$ has much less influence on the inter-system variability of $t_\psi$ than the other factors.

Finally, the dissipation of these equilibrium tides is mainly provided by the eddy viscosity in the convective envelope of the star. This introduces an additional term related to the relative mass of this envelope \citep[following e.g.,][]{Rasio1996,Privitera2016} in Eq.~(\ref{eq:to}), which cannot be neglected as the mass of the convective envelope can arbitrarily approach zero for hot stars. As a result, the tidal efficiency factor used in the present work, proportional to $1/t_\psi$, is given by

\begin{equation}
\label{eq:tau}
  \tau \equiv \frac{M_\mathrm{conv}}{M_\star} \left(\frac{M_\mathrm{p}}{M_\star}\right)^2 \left(\frac{a}{R_\star}\right)^{-6}.
\end{equation}

\noindent Equation (\ref{eq:tau}) is reminiscent of the empirical tidal efficiency factor devised by \citet{Albrecht2012}. The latter authors emphasized that they could not find a theoretical argument for the linear correlation they identified between the convective mass and the spin--orbit angle distribution. However, this dependence seems justified by invoking the relevant timescales (as seen in Eqs.~(\ref{eq:ta}--\ref{eq:tau})) and by the fact that equilibrium tides are very efficiently dissipated by turbulent friction in the external
convective envelopes of stars \citep{Zahn1977,Rasio1996,Privitera2016}.

It is worth noting that \citet{Albrecht2012} also considered tidal dissipation within stellar radiative envelopes for stars hotter than the Kraft break. As pointed out in their work, this approach can bias the results because of the binary decision one has to make to decide whether a star is cold and convective, or hot and radiative. It is particularly problematic for the many stars with an effective temperature consistent with the Kraft break (see Fig.~\ref{fig:Teff_hist}), a caveat we want to avoid. Furthermore, we recall that equilibrium tides are very efficiently dissipated by turbulent friction in the external convective envelopes of stars, while this dissipation is much less efficient in stably stratified radiative layers \citep{Zahn1977}, adding more weight as to why we only consider dissipation of equilibrium tides in convective zones.

Dissipation of dynamical tides, resulting from the excitation of inertial waves in convective stellar layers, can also play a role on top of equilibrium tides. Nevertheless, dynamical tides in convective zones have a significant impact on planetary orbits only for fast-rotating stars during their pre-main sequence \citep{Rao2018}. Dissipation of dynamical tides can also occur in stellar radiative zones as a result of the propagation of internal gravity waves excited at the interface between convective and radiative regions. This can lead to an efficient tidal dissipation, provided that these waves are efficiently damped, as it is the case if wave breaking is able to occur in stellar interiors. However, wave breaking only occurs when the mass of the planet is higher than a given threshold. As shown by \citet{Barker2020}, in the case of low-mass and solar-type stars, this critical planet mass becomes lower than 1 $M_{\rm Jup}$ only very close to the end of the main sequence (a typical value of about 3 $M_{\rm Jup}$ is found in the case of the Sun). Consequently, for all but the most massive planets, tidal dissipation by gravity waves is able to play a dominant role for the orbital evolution only at the end of the main sequence and during the post-main sequence phase \citep{Barker2020}. This explains why we consider in the present work a tidal efficiency factor $\tau \; \propto \; 1/t_\psi$ as given by Eq.~(\ref{eq:tau}).

\begin{table}
\caption{Tabulated values of the mass of the stellar convective envelope, used to compute the tidal efficiency factor, as a function of the stellar mass.}
\label{tab:mconv}
\centering
\begin{tabular}{c c c}
\hline \hline
$M_\star$ [$M_\odot$] & $\log T_{\rm eff}$ [K] & $M_{\rm conv}/M_\star$ \\
\hline
0.5 & 3.60337 & 0.3508 \\
0.6 & 3.62040 & 0.1842 \\
0.7 & 3.64646 & 0.0991 \\
0.8 & 3.68716 & 0.0667 \\
0.9 & 3.72492 & 0.0437 \\
1.0 & 3.75427 & 0.0257 \\
1.1 & 3.77156 & 0.0107 \\
1.2 & 3.79455 & 0.0031 \\
1.3 & 3.81339 & 0.0003 \\
1.4 & 3.83224 & 0.0000 \\
1.5 & 3.85367 & 0.0000 \\
\hline
\end{tabular}
\begin{tablenotes}
\textit{Note:} the stellar effective temperature, as generated by our model, is given for indicative purposes. In particular, one can see that $M_{\rm conv}/M_\star$ drops below 1\% at roughly the temperature of the Kraft break.
\end{tablenotes}
\end{table}


\subsection{Tidal efficiency trend}
\label{sec:tideff_trend}

\begin{figure*}
\includegraphics[width=0.96\textwidth]{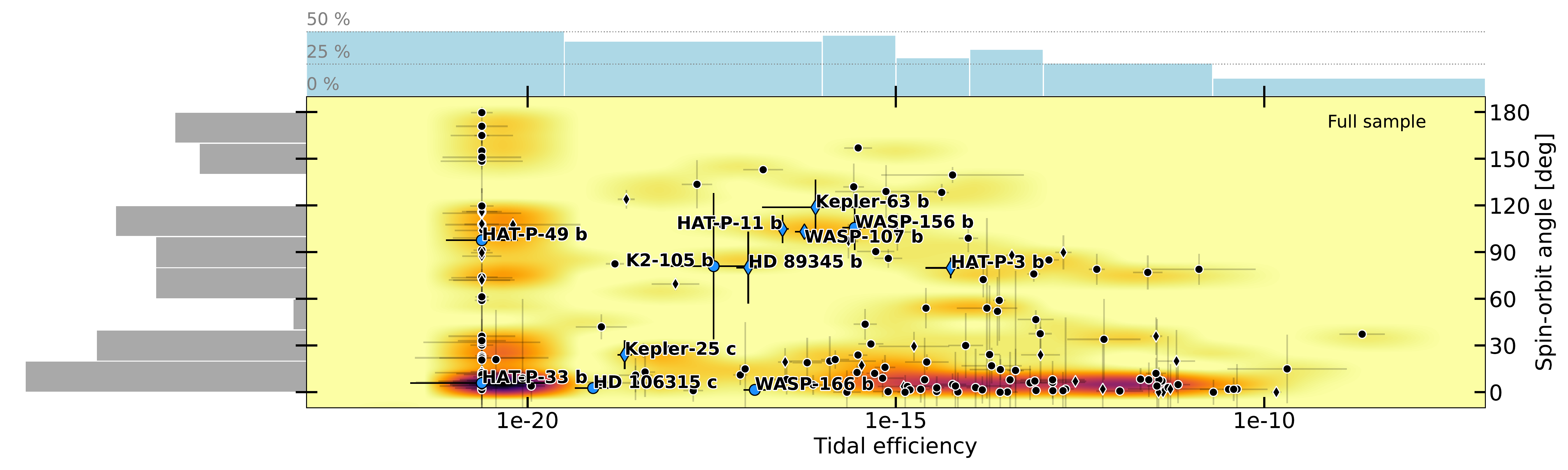}
\centering
\caption{Distribution of spin--orbit angles as a function of the tidal efficiency factor $\tau$. Same color, symbol, and histogram schemes as in Fig.~\ref{fig:Teff_hist}. All systems with a zero tidal efficiency factor have been set to the minimal value of the sample. The gray histogram on the left of the plot covers only the first (lowest) tidal efficiency bin.}
\label{fig:tides_hist}
\end{figure*}

The following step is to compute $\tau$ for all the systems with a known spin--orbit measurement to see if we confirm the trends we previously analyzed. To this effect, $M_{\rm conv}$ is directly obtained from the computation of stellar models for different masses using the Geneva stellar evolution code \texttt{GENEC} \citep{Eggenberger2008}. We provide tabulated values of $M_{\rm conv}$ that can be interpolated for any useful purpose in Table \ref{tab:mconv}. Because $M_{\rm conv}$ and thus $\tau$ can be analytically null, all such systems are set to the minimal nonzero value for $\tau$ within the full sample of investigated planets, so as to still be able to compare orders of magnitude in log-space. The main uncertainty while computing the values of the convective mass comes from the uncertainty on the total stellar mass, since $M_{\rm conv}$ steeply varies as $M_\star$ changes (see Table \ref{tab:mconv}). We hence scale the error on $M_{\rm conv}$ as the error on $M_\star$. This way, we set $\sigma \left(M_{\rm conv}\right)/M_{\rm conv} = \sigma \left(M_\star\right)/M_\star$, which allows us to define error bars on the tidal efficiency factor as well.

The resulting plot is shown in Fig.~\ref{fig:tides_hist}, which shows a downward misalignment trend, indicating that systems with a lower tidal efficiency tend to be more misaligned. Figure \ref{fig:tides_hist} hence combines and confirms the trends seen in Sect.~\ref{sec:trends_tides}, strengthening the role of tides in determining the distribution of known spin--orbit angles. An interesting outlier from our survey is the polar orbit of HAT-P-3 b, which should not have been affected by tidal effects during the main sequence of the star according to \citet{Mancini2018}. However, our own estimates (Fig.~\ref{fig:tides_hist}) show that the system is in the range of tidal efficiency where the orbit could have been realigned. In fact, it is the planet with the highest $\tau$ value in our survey, which suggests that its polar orbit may have been occasioned by a disruptive mechanism recently enough for the realignment process to be still ongoing.

Despite the simplicity of our theoretical tidal framework, our analysis provides evidence that spin--orbit angles are damped by tidal forces and that strong star--planet interactions erase orbital histories by realigning systems. Our tidal efficiency factor could be employed as a useful criterion for future studies of close-in orbital architectures. Based on the trend in the $\lambda - \tau$ bar plot of Fig.~\ref{fig:tides_hist}, we tentatively propose $\tau \sim 10^{-15}$ as a threshold when tidal realignment needs to be considered. 

In this context, we construct a histogram of spin--orbit angles (left of Fig.~\ref{fig:tides_hist}) limited to systems within the lowest tidal efficiency bin, which can be seen as pristine systems unaffected by stellar interactions. We recover the aforementioned aligned and polar populations, and we unveil for the first time a third, distinct group composed of antialigned systems. If confirmed, this surprising distribution is indicative of disruptive processes that happened in the dynamical past of the observed systems, and which were not damped by star--planet interactions. This result suggests that in addition to the ``preponderance of perpendicular orbits'' brought to light by \citet{Albrecht2021}, antialigned orbits unaffected by tides might also represent a noteworthy feature in the global obliquity distribution. Interestingly, one can see in Fig.~\ref{fig:tides_hist} that there are nearly no planets on antialigned orbits outside of the lowest $\tau$ bin. Even though we lack for now more observational evidence to confirm these findings, we can still make two hypotheses based on them. First, that the processes that form retrograde orbits (e.g., Kozai--Lidov resonance following a spike in eccentricity, \citealt{Naoz2011}, or a primordial flip of the protoplanetary disk, \citealt{Zanazzi2018}) do not lead to a uniform spin--orbit angle distribution over $90 - 180^{\circ}$, but preferentially form perpendicular and antialigned orbits. Secondly, that tides quench antialigned orbits back to a perpendicular configuration, or prevent their formation in the first place. A sample of spin--orbit angles covering a broader range of tidal efficiencies, and more detailed studies of processes that can excite spin--orbit angles (e.g., Kozai--Lidov resonance, \citealt{Kozai1962,Lidov1962,Fabrycky2007,Naoz2011}, secular resonance crossing, \citealt{Petrovich2020}, or magnetic warping, \citealt{Foucart2011,Romanova2021}) are needed to investigate our hypothesis. In all cases, systems within this low-$\tau$ bin, like the polar orbit of HAT-P-49 b, are targets of choice for studying the lone impact of secular or primordial dynamical processes since their orbital architectures are representative pictures of these mechanisms, unaltered by tidal effects.


\section{Estimating the true distribution of obliquities}
\label{sec:estim_psi}


\subsection{Reconstructing the obliquity sample}
\label{sec:estim_psi_procedure}

The largest limitation of this analysis is that we lack a measurement of the true 3D spin--orbit angle $\psi$ for a substantial number of planets in the sample. $\psi$ can be disentangled from $\lambda$ if one has knowledge on the orbital $i_{\rm p}$ and stellar $i_\star$ inclinations \citep[e.g.,][]{Winn2007}

\begin{equation}
\label{eq:cospsi}
    \cos \psi = \sin i_\star \sin i_\mathrm{p} \cos \lambda + \cos i_\star \cos i_\mathrm{p},
\end{equation}

\noindent which poses the problem of acquiring a robust measurement of $i_\star$. The most straightforward way to estimate it is assuming stellar solid body rotation \citep[e.g.,][]{borucki_summers1984,Doyle1984}

\begin{equation}
\label{eq:sinis}
    \sin i_\star = \frac{v_\mathrm{eq} \sin i_\star}{v_\mathrm{eq}} = \frac{v_\mathrm{eq} \sin i_\star}{2 \pi R_\star/P_\mathrm{eq}},
\end{equation}

\noindent where $P_{\rm eq}$ is the stellar equatorial rotation period. It is important to stress that directly using Eq.~(\ref{eq:sinis}) to generate an $i_\star$ distribution is hazardous, because $v_{\rm eq} \sin i_\star$ and $P_{\rm eq}$ are correlated quantities. More generally, \citet{Masuda2020} showed that naively sampling a $\psi$ distribution using Eqs.~(\ref{eq:cospsi}, \ref{eq:sinis}) could result in very large biases because of the various correlations between the involved parameters. An accurate determination of $i_\star$ is in reality even more problematic because of the often large uncertainties in the different involved quantities \citep{fabrycky2009,Munoz2018}. 

Keeping these limitations in mind, we seek to recover true spin--orbit angles $\psi$ for all close-in planets in order to assess whether our conclusions stem from sky-projection effects or not. Regarding the 154 sky-projected measurements in the sample, we need knowledge on the stellar inclination to disentangle $\psi$ from $\lambda$ with Eq.~(\ref{eq:cospsi}). For the 28 systems of this subsample that have a $P_{\rm eq}$ measurement, we propose the following procedure to derive $i_\star$, so as to avoid the aforementioned biases outlined in \citet{Masuda2020}. We fit for $v_{\rm eq} \sin i_\star$ with a Markov chain Monte Carlo \citep[MCMC, using the \texttt{emcee} package,][]{Foreman2013} based on the following formulation of Eq.~(\ref{eq:sinis}) 

\begin{equation}
\label{eq:vsini}
    v_\mathrm{eq} \sin i_\star = \frac{2 \pi R_\star}{P_\mathrm{eq}} \sqrt{1 - \cos^2 i_\star},
\end{equation}

\noindent and using the literature value as a constraint for the computation of the likelihood. We set $R_\star$, $P_{\rm eq}$, and $\cos i_\star$ as independent jump parameters with measurement-informed Gaussian priors on $R_\star$ and $P_{\rm eq}$, and a uniform prior on $\cos i_\star$ (i.e., assuming an isotropic stellar inclination distribution). We then derive from the results the probability distribution function (PDF) of the stellar inclination, which we use to generate $\psi$ using Eq.~(\ref{eq:cospsi}). The two degenerate configurations corresponding to $i_\star$ and $\pi - i_\star$, assumed to be equiprobable, are combined to yield the final PDF of $\psi$. Using this procedure, based on Eq.~(\ref{eq:vsini}), circumvents the interdependence problem that \citet{Masuda2020} warned about. Finally, for each of the remaining 126 systems, we sample $\psi$ through Eq.~(\ref{eq:cospsi}), using the observational Gaussian distributions of $\lambda$ and $i_{\rm p}$, and a uniform distribution for $\cos i_\star$. For all the systems for which we reconstruct the 3D obliquity, we retain the median of the resulting distribution as an estimate for $\psi$ and the 68\% highest density interval (HDI) to define the error bars.

We include in App.~\ref{app:sample} the values of $\psi$ derived by fitting for $v_{\rm eq} \sin i_\star$, as they can be readily used as accurate 3D spin--orbit angles for the concerned systems. We choose not to include the values of $\psi$ derived only using the isotropic stellar inclination assumption, as we deem that this approach should not be employed for individual systems, but for statistical studies like this one.


\subsection{Analyzing the true distribution of obliquities}
\label{sec:distrib_psi}

\begin{figure}
\includegraphics[width=0.45\textwidth]{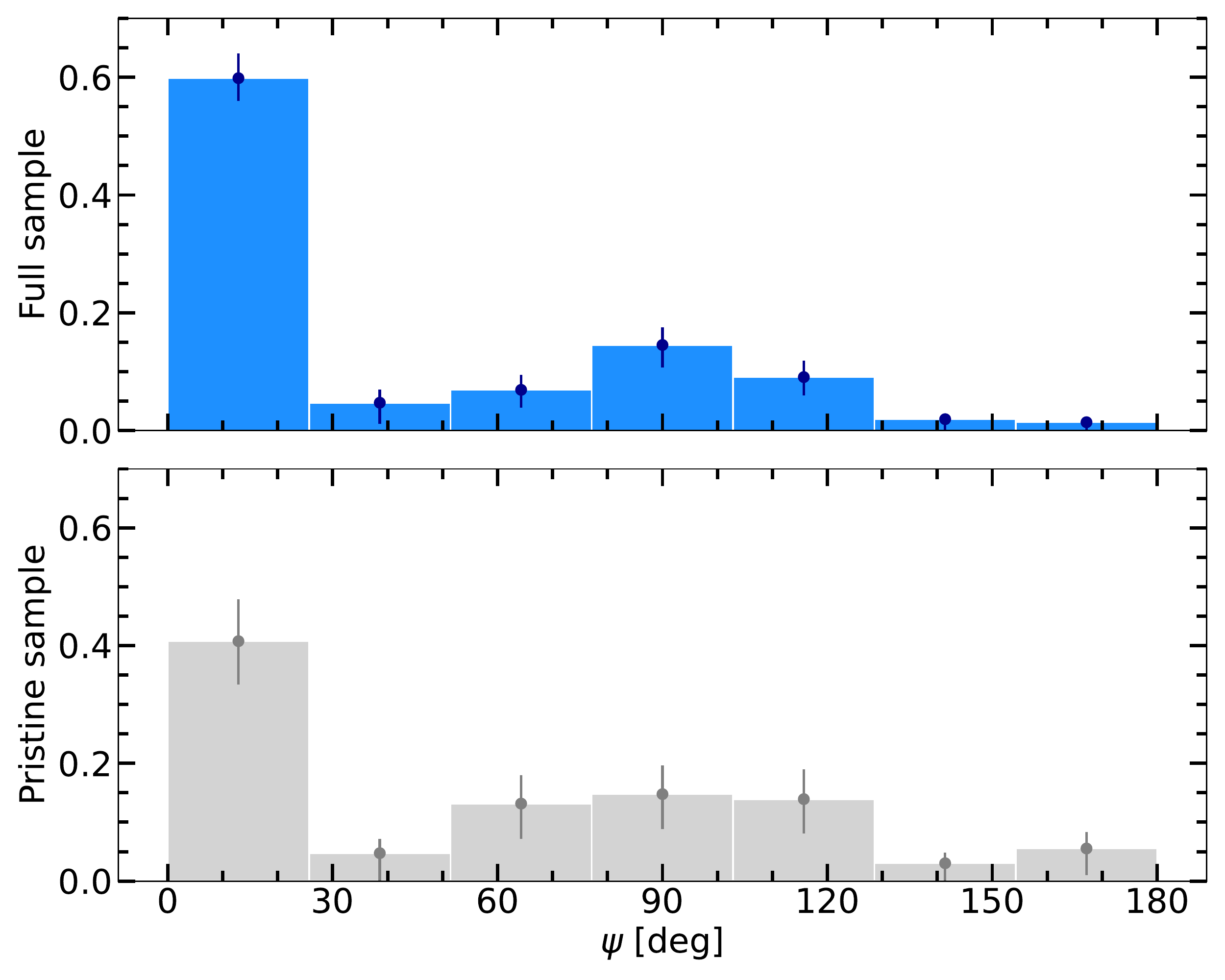}
\centering
\caption{Histogram of 3D spin--orbit angles for the full (top) and pristine (bottom) samples. The $\psi$ distributions for individual systems are estimated using the procedure outlined in Sect.~\ref{sec:estim_psi_procedure} and the histograms are built following the approach of Sect.~\ref{sec:distrib_psi}.}
\label{fig:psi_hist}
\end{figure}

Following the approach of Sect.~\ref{sec:estim_psi_procedure}, we are able to estimate the 3D, true, spin--orbit angles for all systems with an obliquity measurement, which calls for analyzing their global distribution to see if ensemble features emerge. To this end, it would be tempting to build an obliquity histogram by naively counting the number of systems in each spin--orbit angle bin, as in Fig.~\ref{fig:Teff_hist} (top panel, right of the figure). However, constructing such a histogram solely based on the medians of the estimated $\psi$ distributions would lead to large biases in the final results, since it does not capture the uncertainty relative to falling into one of two adjacent bins. Plus, the computed $\psi$ distributions are not necessarily Gaussian and might be skewed, making the median an unreliable estimator for such a procedure.

In order to construct a statistically more robust histogram that is informed by the full PDFs of the $\psi$ distributions, and not just their medians, we design a Bayesian framework with an MCMC implementation, setting the jump parameters as the consecutive $\boldsymbol{\theta} = \left( \theta_1, ..., \theta_K \right)$, which represent the fraction of systems that fall in each of the $K$ obliquity bins. As these fractions need to be positive and satisfy $\sum_{i = 1}^K \theta_i= 1$, we impose a Dirichlet prior on their joint distribution, with a unitary parameter in each dimension (i.e., a flat Dirichlet distribution). This can be seen as a uniform, noninformative, prior over the open $K - 1$ simplex. The computation of the likelihood of the data $\mathcal{D}^j$ for one system $j$ can be derived using a hierarchichal model

\begin{equation}
    p(\mathcal{D}^j \; | \; \boldsymbol{\theta}) = \int p(\mathcal{D}^j \; | \; \psi^j) p(\psi^j \; | \; \boldsymbol{\theta}) \; \mathrm{d}\psi^j,
\end{equation}

\noindent where $p(\mathcal{D}^j \; | \; \psi^j)$ can be drawn from the expression of the posterior probability distribution of the spin--orbit angle for that system $j$

\begin{equation}
    p(\psi^j \; | \; \mathcal{D}^j) \; \propto \; p(\mathcal{D}^j \; | \; \psi^j) \; p(\psi^j),
\end{equation}

\noindent with $p(\psi^j)$ being the prior function on $\psi^j$, which leads to

\begin{equation}
    p(\mathcal{D}^j \; | \; \boldsymbol{\theta}) \; \propto \; \int p(\psi^j \; | \; \mathcal{D}^j) \frac{p(\psi^j \; | \; \boldsymbol{\theta})}{p(\psi^j)} \; \mathrm{d}\psi^j.
\end{equation}

\noindent The previous equation can be computed by means of a Monte Carlo integration, based on importance sampling

\begin{equation}
\label{eq:like}
    p(\mathcal{D}^j \; | \; \boldsymbol{\theta}) \; \appropto \; \sum_i \frac{p(\psi_i^j \; | \; \boldsymbol{\theta})}{p(\psi_i^j)},
\end{equation}

\noindent where the various $\psi_i^j$ are sampled according to their previously generated posterior distribution (Sect.~\ref{sec:estim_psi_procedure}) for the system $j$. The terms of Eq.~(\ref{eq:like}) can be easily evaluated, as $p(\psi_i^j \; | \; \boldsymbol{\theta}) = \theta_k$ if $\psi_i^j$ falls into the obliquity bin $k$, and $p(\psi_i^j) \; \propto \; \sin \psi_i$. The latter point comes from an examination of the employed prior to derive the $\psi$ posteriors, namely a uniform prior on $\cos \psi$. It can be seen as a manifestation of the isotropic assumption, that we recover with Eq.~(\ref{eq:cospsi}) by setting isotropic distributions for the other angles, that is to say uniform on $\cos i_{\rm p}$, $\cos i_\star$, and $\lambda$ (and not $\cos \lambda$, as it is the projection of an angle on a plane). Since the obliquity values of different systems are independent from each other, the total log-likelihood $p(\mathcal{D} \; | \; \boldsymbol{\theta})$ for one set of $\boldsymbol{\theta}$ is the sum of the log-likelihoods for the individual systems, which are computed using Eq.~(\ref{eq:like}).

Again, we make use of \texttt{emcee} to set up the MCMC simulation, adjusting the number of walkers and the burn-in phase according to the convergence of the chains. Figure~\ref{fig:psi_hist} shows the results, the top panel featuring the entire sample, and the bottom one being limited to systems within the lowest tidal efficiency bin (as in Sect.~\ref{sec:tideff_trend}). The height of the bars is set by the medians of the $\theta_i$ marginal PDFs, and the error bars are defined by their 68\% HDIs.

The first striking result is that our reconstruction of the 3D obliquity histogram for the whole population (Fig.~\ref{fig:psi_hist}, top panel) showcases an significant peak around $90^\circ$. The preponderance of perpendicular orbits spotlighted by \citet{Albrecht2021} seems to be a robust feature, especially considering the fact that our sample is nearly four times larger. It is worth emphasizing that this prominence of perpendicular systems cannot be an artifact of our choice of prior function on $\psi$, as polar architectures are actually penalized by the likelihood function (Eq.~(\ref{eq:like})). It is also remarkable that the fraction of polar orbits within the entire population on the one hand, and the pristine population (Fig.~\ref{fig:psi_hist}, bottom panel) on the other hand, is roughly equivalent ($\sim 15$ \%). This noteworthy feature suggests that the mechanisms generating polar systems, either primordial or evolutionary, are resilient against tidal realignment. These mechanisms could be the result of spin--orbit resonances leading to such orbits in a dynamically stable configuration.

Contrastingly, the fraction of aligned systems is significantly different between the two samples ($\sim 3 \sigma$), stressing again the role of tidal effects in realigning planetary systems. The substantially larger number of truly aligned systems compared to the low-efficiency distribution implies that many systems end up aligned as a result of tidal interactions rather than through formation processes. In fact, the relatively small fraction of aligned orbits in a distribution of pristine systems unaffected by star--planet interactions ($\sim 40$ \%) somewhat reverses the classical picture that the majority of planets smoothly form in the disk, keeping alignments intact. This classical picture is further challenged by the very existence of antialigned orbits (that might surprisingly be as numerous as 10 \% of the pristine sample, within $1 \sigma$). The absence of flipped orbits in the entire population indicates that tidal interactions play a definite role in quenching antialigned architectures.


\subsection{Reassessing the tidal efficiency trend}
\label{sec:tau_psi}

\begin{figure*}
\includegraphics[width=0.96\textwidth]{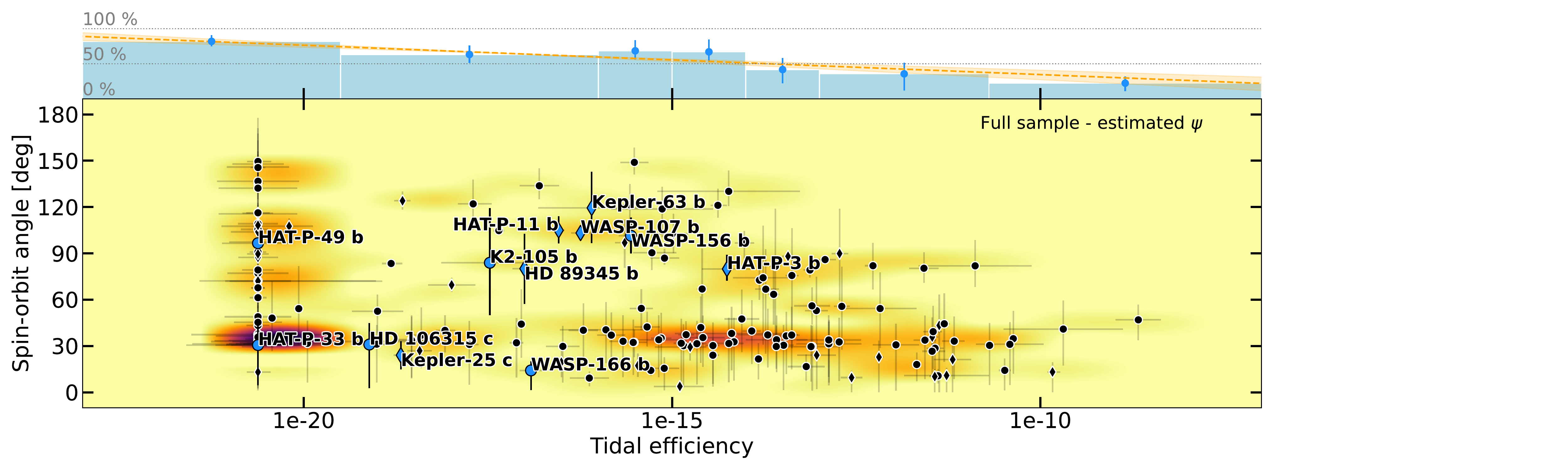}
\centering
\caption{Same as Fig.~\ref{fig:tides_hist}, but with 3D obliquities for all systems. $\psi$ is estimated for systems that only have a sky-projected measurement using the procedure outlined in Sect.~\ref{sec:estim_psi_procedure}. We note the different vertical range for the blue bar plot on the top of the figure as compared to Fig.~\ref{fig:tides_hist}. The bar plot is built according to the procedure described in Sect.~\ref{sec:tau_psi}. The orange dashed line on the bar plot is the best linear fit between the misalignment fraction and the tidal efficiency, with the orange shaded area around it shown as its $1 \sigma$ envelope. The bar plot, along with its linear fit, is also separately shown in Fig.~\ref{fig:theta_tau} for clarity.}
\label{fig:tides_hist_psi}
\end{figure*}

\begin{figure}
\includegraphics[width=0.45\textwidth]{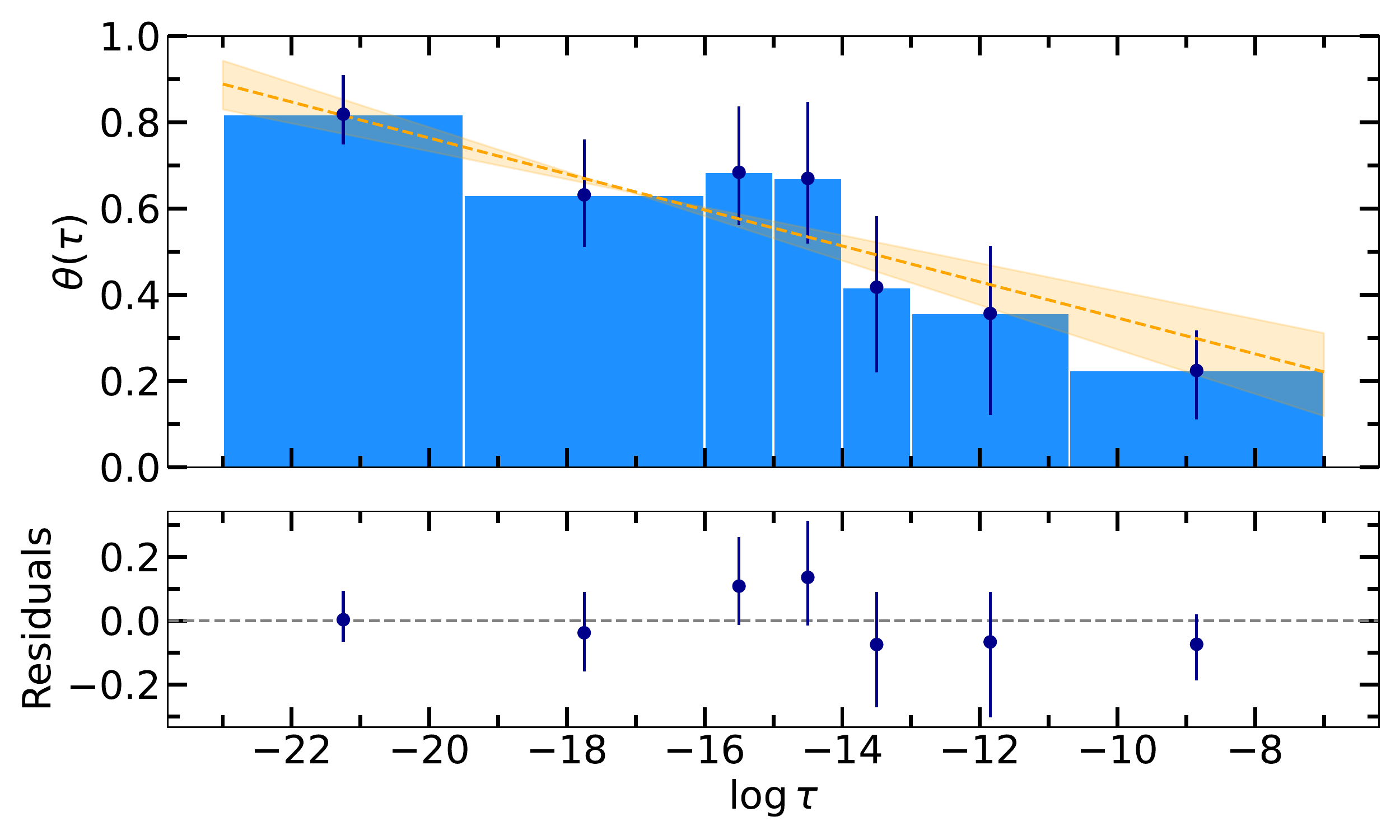}
\centering
\caption{Misalignment fraction as a function of the tidal efficiency factor for all systems in our sample. \textit{Top:} misalignment fraction bar plot, reported from the top of Fig.~\ref{fig:tides_hist_psi}. The best linear fit relation (Eq.~(\ref{eq:misrate_tau})) can be used as an estimator for the probability that a system lies in a misaligned configuration. \textit{Bottom:} residuals between the medians of the misalignment fractions and the linear fit.}
\label{fig:theta_tau}
\end{figure}

\begin{figure}
\includegraphics[width=0.45\textwidth]{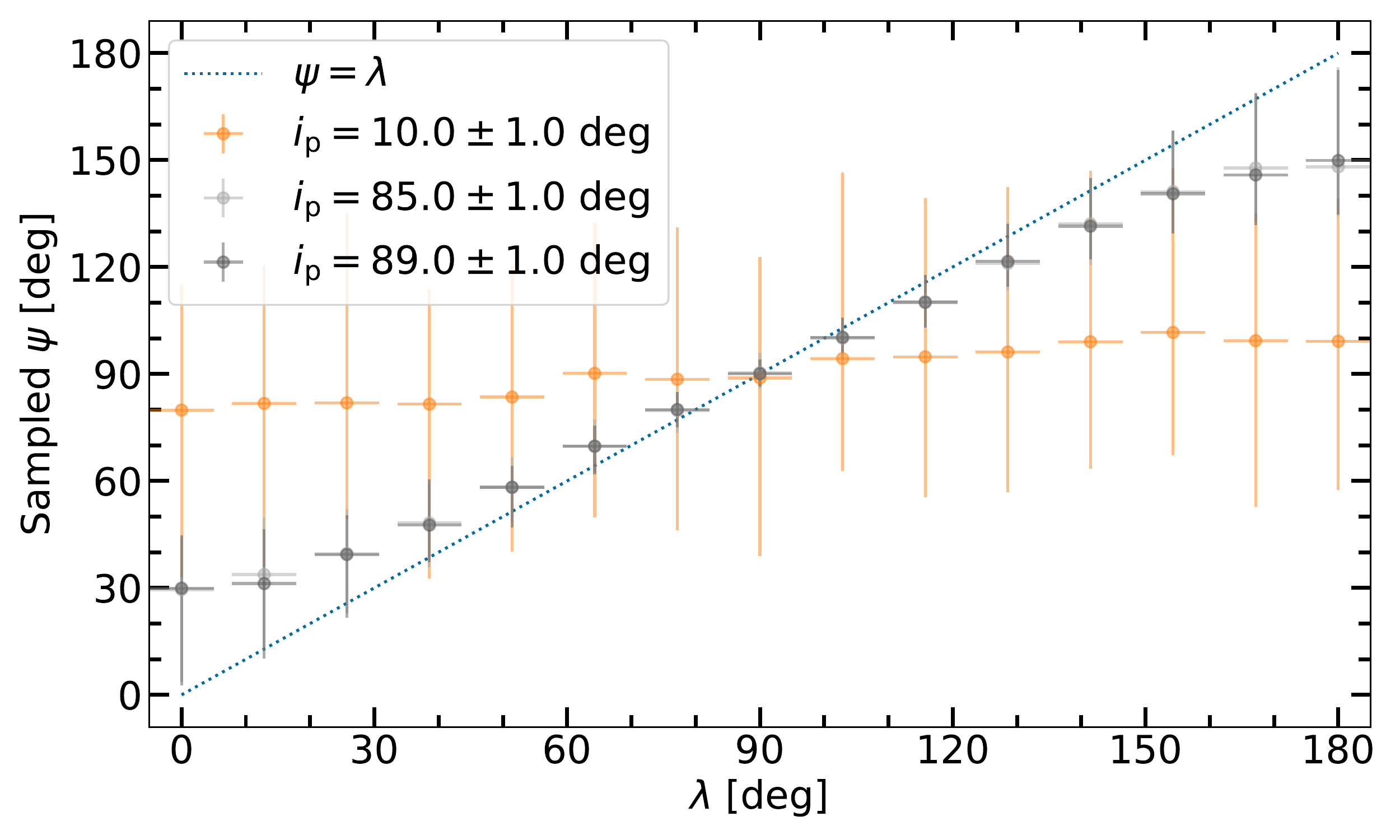}
\centering
\caption{Estimates of $\psi$ as a function of $\lambda$ under the isotropic stellar inclination assumption, for three different values of the orbital inclination. Distributions for $\psi$ are computed as described in the main text, assuming Gaussian distributions for $\lambda$ and $i_{\rm p}$ with respective standard deviations of $5^\circ$ and $1^\circ$. Plotted values are the medians of the $\psi$ distributions, with error bars set to their 68\% HDIs.}
\label{fig:psi_sample}
\end{figure}

Equipped with the estimated $\psi$ distributions (Sect.~\ref{sec:estim_psi_procedure}), we reconstruct in Fig.~\ref{fig:tides_hist_psi} the tidal efficiency plot using 3D obliquities. In particular, we reanalyze the fraction of misaligned systems (bar plot, top of Fig.~\ref{fig:tides_hist_psi}, also separately shown in Fig.~\ref{fig:theta_tau} for clarity), which we determine this time using the same procedure as for constructing the histogram in Fig.~\ref{fig:psi_hist}. We run a joint MCMC simulation for all of the $\tau$ bins at once, the jump parameters being the fraction of misaligned systems within each bin. In addition to using the information of the full $\psi$ PDFs and alleviate the possible biases resulting from a naive count of misaligned systems based on the $\psi$ medians alone, this procedure takes into account the error bars on the $\tau$ values as well, making our analysis more robust against the potential uncertainty relative to falling into one of two adjacent tidal efficiency bins. The heights of the blue bars are hence defined by the medians of the misalignment fraction PDFs, and their 68\% HDIs set the error bars.

First, we see that the number of misaligned systems is substantially higher in all $\tau$ bins as compared to Fig.~\ref{fig:tides_hist}. This can be understood by examining our procedure to estimate the true spin--orbit angle under the assumption of isotropic stellar inclination, which may have revealed an important observational bias. In Fig.~\ref{fig:psi_sample}, we show values of $\psi$ estimated using this approach as a function of $\lambda$, for three different values of $i_{\rm p}$. The true and sky-projected spin--orbit angles are similar around 90$^\circ$, but the estimated $\psi$ deviates from $\lambda$ when the latter goes toward lower or higher values, up to a bias of about 90$^\circ$ for $\lambda \sim 0^\circ$ or $180^\circ$. This explains the global increase in misaligned systems in Fig.~\ref{fig:tides_hist_psi}, as systems considered as ``aligned'' based on their sky-projected spin--orbit angle ($\lambda < 30^\circ$ in Fig.~\ref{fig:tides_hist}) are more likely to be misaligned based on their estimated true spin--orbit angle (where $\psi \gtrsim 30^\circ$ for transiting systems, Fig~\ref{fig:psi_sample}). This is exactly the case for HAT-P-33 b, for which the sky-projected obliquity $\lambda$ could be interpreted as a surprisingly aligned orbit (Fig.~\ref{fig:tides_hist}) despite the shallow stellar convective envelope (Fig.~\ref{fig:Teff_hist}), but may be more expectedly misaligned in reality (Fig.~\ref{fig:tides_hist_psi}). This bias is actually even more accentuated for nontransiting systems ($i_{\rm p} = 10^\circ$ in Fig.~\ref{fig:psi_sample}), for which the true spin--orbit angle is almost always estimated as polar whatever the value of $\lambda$. In a sense, Fig.~\ref{fig:psi_sample} shows that all sky-projected methods for determining the spin--orbit angle might be biased toward misaligned systems, even the ones not limited to transiting systems \citep[e.g., interferometry,][]{Kraus2022}. 

Another important feature in Fig.~\ref{fig:tides_hist_psi} is the very high occurence rate of misaligned orbits in the DREAM I sample. Noticeably, two thirds of the surveyed systems are consistent with a polar architecture within $< 2 \sigma$, strongly contrasting with the $\sim 15\%$ fraction of polar orbits we find for the entire population of exoplanets with an obliquity measurement (Fig.~\ref{fig:psi_hist}). Even though the latter value should be considered with caution because of the various selection biases (e.g., hot Jupiters are by far the most represented category of planets with a characterized spin--orbit angle, Fig.~\ref{fig:population}), this dichotomy might be indicative of disruptive dynamical processes that particularly affects planets spanning the edge of the desert and the breadth of the savanna. Consequently, high-eccentricity migration scenarios, which naturally lead to highly misaligned configurations, may be the predominant process to bring these planets from their birthplace to their current close-in orbit.

\begin{figure}
\includegraphics[width=0.45\textwidth]{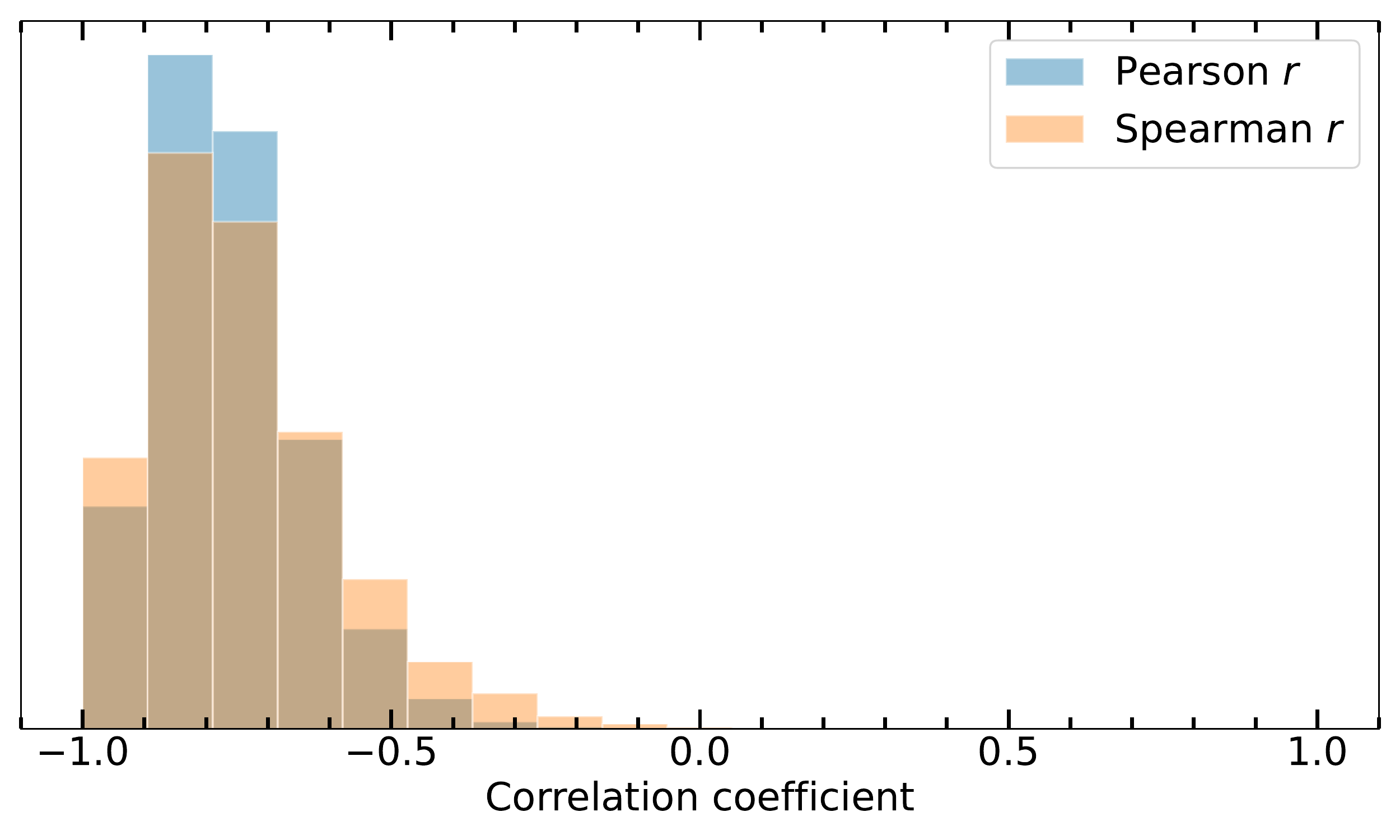}
\centering
\caption{Distribution of Pearson (blue) and Spearman (orange) $r$ correlation coefficients between misalignment fractions (blue bars on top of Fig.~\ref{fig:tides_hist_psi}) and log-tidal efficiencies (Eq.~(\ref{eq:tau})). Misalignment fractions are randomly sampled from their PDFs to generate this distribution (Sect.~\ref{sec:tau_psi}).}
\label{fig:corr_coeff}
\end{figure}

Furthermore, we notice that the downward misalignment trend with increasing tidal efficiencies identified in Sect.~\ref{sec:tideff_trend} persists. Motivated by the visible robustness of this result, we quantify this trend using the Spearman and Pearson $r$ statistical tests, which respectively measure rank and linear correlations. To this end, misalignment fractions are randomly sampled from their PDFs and used to compute a distribution of Spearman and Pearson coefficients. The results are presented in Fig.~\ref{fig:corr_coeff}, which shows a clear anticorrelation for both coefficients. The misalignment fraction not only decreases monotonically with an increasing log-tidal efficiency (Spearman $r = -0.79^{+0.07}_{-0.18}$), but the two quantities are moreover linearly anticorrelated (Pearson $r = -0.78^{+0.09}_{-0.13}$). The two coefficients are in fact consistent with $-1$ (perfect anticorrelation) within $< 2 \sigma$. Defining the null hypothesis as the absence of, or a positive correlation ($r \geqslant 0$), the respective $p$-values are $0.080\%$ and $0.003\%$, highlighting again the robustness of the result. The best linear fit between the misalignment fraction and the tidal efficiency factor, along with its $1 \sigma$ uncertainty envelope, is shown on top of Fig.~\ref{fig:tides_hist_psi}, for which we provide here an explicit expression

\begin{equation}
\label{eq:misrate_tau}
    \theta \left( \tau \right) = \alpha \times \log_{10} \left(\tau\right) + \beta,
\end{equation}
\begin{equation*}
    \alpha = -0.0417^{+0.0093}_{-0.0098}, \; \beta = -0.0706^{+0.1550}_{-0.1712},
\end{equation*}

\noindent $\theta$ being the misalignment fraction. The $\theta \left( \tau \right)$ bar plot, as well as its linear fit, can be seen in isolation in Fig.~\ref{fig:theta_tau}, which further shows residuals satisfactorily consistent with zero within $1 \sigma$ for all tidal efficiency bins. Equation (\ref{eq:misrate_tau}) can be conveniently used for future studies, as $\theta \left( \tau \right)$ can be interpreted as the probability that a system with a tidal efficiency factor $\tau$ is misaligned. We note that this relationship is satisfyingly consistent with the threshold we set for considering tidal realignment processes (Sect.~\ref{sec:tideff_trend}), since $\theta \left( \tau = 10^{-15} \right) \simeq 50\%$. 

We highlight that the conclusions we put forward in this section do not depend on the threshold of $30^\circ$ we set to define misaligned orbits. By conducting the same analysis for thresholds of $20^\circ$ and then $10^\circ$, we expectedly see increasingly higher misalignment fractions within all the tidal efficiency bins. The downward misalignment trend remains nevertheless intact, with Spearman and Pearson $r$ coefficients still consistent with a perfect anticorrelation within $\sim 1.5 \sigma$ and slightly higher $p$-values (yet below 0.5\%). The only substantial change while considering a different misalignment threshold is the value of $\beta$ in Eq.~(\ref{eq:misrate_tau}), but not its slope. The values of $\alpha$ corresponding to the different thresholds are in fact consistent with each other within their error bars. We thus recommend keeping in mind our conservative definition of a misaligned orbit in future works, should Eq.~(\ref{eq:misrate_tau}) be employed.

\section{Summary and conclusion}
\label{sec:conclu}

In this work, we conducted a global analysis of the distribution of known spin--orbit angles. We collected a reliable database of exoplanet parameters after a critical review of the sources, and complemented it with the obliquity measurements derived in DREAM I \citep{Bourrier2023}. The resulting large sample of 196 spin--orbit angles (App.~\ref{app:sample}) allowed us to draw a detailed picture of the orbital architectures of close-in exoplanets. We confirmed the strong correlation between misalignment fractions and the weakness of star--planet tidal interactions by refining trends with individual parameters controlling the intensity of tides like the stellar effective temperature, the planet-to-star mass ratio, or the scaled separation, in line with \citet{Triaud2018} and \citet{Albrecht2022}, for example.

Going a step further, we devised a tidal efficiency parameter $\tau$ combining several pertinent planetary, stellar, and orbital quantities \citep[Eq.~(\ref{eq:tau}), inspired by][]{Albrecht2012} and made use of it to corroborate our conclusions. While the aforementioned trends with individual parameters can only be seen by examining relevant sub-populations in isolation, the $\lambda - \tau$ correlation is explicit for the whole exoplanet sample irrespective of their individual specificities: systems where tidal forces are weaker tend to be more misaligned and vice-versa, strongly suggesting that tidal realignment drives the global distribution of spin--orbit angles. We can thus propose for future studies the use of this tidal efficiency parameter as a criterion to assess the impact of tidal realignment, calculating it based on the explicit recipe we lay out (Eq.~(\ref{eq:tau}), Table \ref{tab:mconv}).

Being aware that our conclusions might be biased by sky-projection effects, we reconstructed the distribution of true, 3D, obliquities, under the assumption of isotropic stellar inclinations. In doing so, the anticorrelation of misalignment fractions $\theta$ with increasing tidal efficiencies $\tau$ persists, proving the robustness of this result. Indeed, statistical tests strongly support a linear anticorrelation between the two quantities, and we provide an explicit formula for the best fit ($\theta \left( \tau \right)$, Eq.~(\ref{eq:misrate_tau})). This expression can be conveniently used for estimating the probability that a system is misaligned, which will prove useful in particular for setting the distribution of misalignments in global population studies of close-in exoplanets.

We also showed that misaligned orbits do not randomly span the entire possible range of obliquities. Rather, the distribution of $\psi$ favors polar architectures, a result first identified by \citet{Albrecht2021}, which we confirm with a sample nearly four times larger. On the other hand, pristine systems that do not feel tidal interactions may showcase a flatter distribution around $90^\circ$ than the entire population, implying that polar orbits are a stable outcome of dynamical processes that can hardly be aligned back by tides.

Our results suggest that exoplanet systems may be born with a broad, non-unifrom, range of spin--orbit angles, only to be realigned thanks to tides if the physical properties of the system allow it. This interpretation is substantiated by comparing the obliquity distribution for all systems and for the pristine ones unaffected by tides, the former showing a stronger preference for aligned orbits. It is all the more interesting to put these conclusions in perspective of the Neptune desert, for which evidence has been put forward that it may be shaped by secular disruptive processes \citep[e.g.,][]{Attia2021}. Such mechanisms, like the Kozai--Lidov resonance, naturally produce highly misaligned orbits that can only be damped if tidal forces are strong enough. In fact, the surveyed sample presented in DREAM I, cherry-picked to span the rims of the desert and the breadth of the savanna, showcases a high occurence rate of considerably misaligned orbits, polar or retrograde for most of them, further hinting that this class of planets may be particularly affected by these disruptive processes. Additional observational efforts on these exciting regions of the parameter space will be pivotal to shed light on the possible history of the planets populating them. Besides expanding the current obliquity sample, we recommend parallel searches for outer companions that could have induced high-eccentricity migration, using both radial velocities and direct imaging.


\begin{acknowledgements}
We thank A.H.M.J.~Triaud for bringing the question of sky-projection effects into our mind and inciting us to look deeper into this matter. This work has been carried out within the framework of the NCCR PlanetS supported by the Swiss National Science Foundation under grants 51NF40$\_$182901 and 51NF40$\_$205606. This project has received funding from the European Research Council (ERC) under the European Union's Horizon 2020 research and innovation programme (project {\sc Spice Dune}, grant agreement No 947634; grant agreement No 730890). This material reflects only the authors' views and the Commission is not liable for any use that may be made of the information contained therein.
\end{acknowledgements}


\bibliographystyle{aa} 
\bibliography{biblio} 


\begin{appendix}

\section{Spin--orbit angles used in this work}
\label{app:sample}

\begin{table*}
\caption{Spin--orbit angle values (projected in the sky plane and 3D) used in our analysis along with their sources.}
\centering
\tiny
\begin{tabular}{ccccc}
\hline \hline
  Planet name &           \multicolumn{2}{c}{Projected spin--orbit angle}  &         \multicolumn{2}{c}{3D spin--orbit angle} \\
              &              $\lambda$ (deg) &                      Source &            $\psi$ (deg) &                 Source \\
\hline
     55 Cnc e &       $72.4^{+12.7}_{-11.5}$ &       \citet{bourrier2014b} &  $72.7^{+12.2}_{-12.7}$ &              This work \\
     AU Mic b &         $-4.7^{+6.8}_{-6.4}$ &          \citet{Hirano2020} &     $9.2^{+4.3}_{-5.3}$ &              This work \\
    CoRoT-1 b &       $77.0^{+11.0}_{-11.0}$ &            \citet{Pont2010} &                     --- &                    --- \\
   CoRoT-11 b &          $0.1^{+2.6}_{-2.6}$ &        \citet{Gandolfi2012} &                     --- &                    --- \\
   CoRoT-18 b &      $-10.0^{+20.0}_{-20.0}$ &         \citet{Hebrard2011} &  $20.0^{+20.0}_{-20.0}$ &    \citet{Hebrard2011} \\
   CoRoT-19 b &      $-52.0^{+27.0}_{-22.0}$ &        \citet{Guenther2012} &                     --- &                    --- \\
    CoRoT-2 b &          $7.2^{+4.5}_{-4.5}$ &          \citet{Bouchy2008} &    $10.7^{+4.3}_{-5.1}$ &              This work \\
    CoRoT-3 b &      $-37.6^{+22.3}_{-10.0}$ &          \citet{Triaud2009} &                     --- &                    --- \\
     DS Tuc b &          $2.9^{+0.9}_{-0.9}$ &            \citet{Zhou2020} &    $14.2^{+2.9}_{-3.6}$ &              This work \\
    GJ 3470 b &      $101.0^{+29.0}_{-14.0}$ &      \citet{Stefansson2022} &  $97.0^{+16.0}_{-11.0}$ & \citet{Stefansson2022} \\
     GJ 436 b &      $114.0^{+23.0}_{-17.0}$ &        \citet{Bourrier2022} & $103.0^{+13.0}_{-12.0}$ &   \citet{Bourrier2022} \\
    HAT-P-1 b &          $3.7^{+2.1}_{-2.1}$ &         \citet{Johnson2008} &                     --- &                    --- \\
   HAT-P-11 b &        $133.9^{+7.1}_{-8.3}$ &        \citet{Bourrier2023} &   $104.9^{+8.6}_{-9.1}$ &   \citet{Bourrier2023} \\
   HAT-P-12 b &      $-54.0^{+41.0}_{-13.0}$ &         \citet{Mancini2018} &                     --- &                    --- \\
   HAT-P-13 b &          $1.9^{+8.6}_{-8.6}$ &           \citet{Winn2010c} &                     --- &                    --- \\
   HAT-P-14 b &        $189.1^{+5.1}_{-5.1}$ &            \citet{Winn2011} &                     --- &                    --- \\
   HAT-P-16 b &       $-2.0^{+55.0}_{-46.0}$ &        \citet{Albrecht2012} &                     --- &                    --- \\
   HAT-P-17 b &       $19.0^{+14.0}_{-16.0}$ &          \citet{Fulton2013} &                     --- &                    --- \\
   HAT-P-18 b &      $132.0^{+15.0}_{-15.0}$ &        \citet{Esposito2014} &                     --- &                    --- \\
    HAT-P-2 b &        $0.2^{+12.2}_{-12.5}$ &        \citet{Loeillet2008} &                     --- &                    --- \\
   HAT-P-20 b &         $-8.0^{+6.9}_{-6.9}$ &        \citet{Esposito2017} &  $36.0^{+10.0}_{-12.0}$ &   \citet{Esposito2017} \\
   HAT-P-22 b &         $-2.1^{+3.0}_{-3.0}$ &         \citet{Mancini2018} &  $24.0^{+18.0}_{-18.0}$ &    \citet{Mancini2018} \\
   HAT-P-23 b &       $15.0^{+22.0}_{-22.0}$ &          \citet{Moutou2011} &                     --- &                    --- \\
   HAT-P-24 b &       $20.0^{+16.0}_{-16.0}$ &        \citet{Albrecht2012} &                     --- &                    --- \\
   HAT-P-27 b &       $24.2^{+76.0}_{-44.5}$ &           \citet{Brown2012} &                     --- &                    --- \\
    HAT-P-3 b &      $-25.3^{+29.4}_{-22.8}$ &        \citet{Bourrier2023} &    $75.7^{+8.5}_{-7.9}$ &   \citet{Bourrier2023} \\
   HAT-P-30 b &         $73.5^{+9.0}_{-9.0}$ &         \citet{Johnson2011} &                     --- &                    --- \\
   HAT-P-32 b &         $85.0^{+1.5}_{-1.5}$ &        \citet{Albrecht2012} &                     --- &                    --- \\
   HAT-P-33 b &         $-5.9^{+4.1}_{-4.1}$ &        \citet{Bourrier2023} &                     --- &                    --- \\
   HAT-P-34 b &        $0.0^{+14.0}_{-14.0}$ &        \citet{Albrecht2012} &                     --- &                    --- \\
   HAT-P-36 b &      $-14.0^{+18.0}_{-18.0}$ &         \citet{Mancini2015} &   $0.0^{+63.0}_{-63.0}$ &    \citet{Mancini2015} \\
    HAT-P-4 b &       $-4.9^{+11.9}_{-11.9}$ &            \citet{Winn2011} &                     --- &                    --- \\
   HAT-P-41 b &        $-22.1^{+0.8}_{-6.0}$ &         \citet{Johnson2017} &                     --- &                    --- \\
   HAT-P-49 b &        $-97.7^{+1.8}_{-1.8}$ &        \citet{Bourrier2023} &                     --- &                    --- \\
   HAT-P-56 b &          $8.0^{+2.0}_{-2.0}$ &            \citet{Zhou2016} &                     --- &                    --- \\
    HAT-P-6 b &        $165.0^{+6.0}_{-6.0}$ &        \citet{Albrecht2012} &                     --- &                    --- \\
   HAT-P-69 b &         $30.3^{+6.1}_{-7.3}$ &            \citet{Zhou2019} &                     --- &                    --- \\
    HAT-P-7 b &        $220.3^{+8.2}_{-9.3}$ &         \citet{Benomar2014} & $115.0^{+19.0}_{-16.0}$ &    \citet{Benomar2014} \\
   HAT-P-70 b &        $107.9^{+2.0}_{-1.7}$ &      \citet{BelloArufe2022} &                     --- &                    --- \\
    HAT-P-8 b &       $-17.0^{+9.2}_{-11.5}$ &          \citet{Moutou2011} &                     --- &                    --- \\
    HAT-P-9 b &        $-16.0^{+8.0}_{-8.0}$ &          \citet{Moutou2011} &                     --- &                    --- \\
    HATS-14 b &         $76.0^{+4.0}_{-5.0}$ &            \citet{Zhou2015} &                     --- &                    --- \\
     HATS-2 b &          $8.0^{+8.0}_{-8.0}$ &   \citet{MohlerFischer2013} &  $28.5^{+14.0}_{-18.0}$ &              This work \\
     HATS-3 b &        $3.0^{+25.0}_{-25.0}$ &         \citet{Addison2014} &                     --- &                    --- \\
    HATS-70 b &          $8.9^{+5.6}_{-4.5}$ &           \citet{Zhou2019b} &    $13.2^{+6.4}_{-5.9}$ &      \citet{Zhou2019b} \\
  HD 106315 c &         $-2.7^{+2.7}_{-2.6}$ &        \citet{Bourrier2023} &                     --- &                    --- \\
  HD 149026 b &         $12.0^{+7.0}_{-7.0}$ &        \citet{Albrecht2012} &                     --- &                    --- \\
   HD 17156 b &         $10.0^{+5.1}_{-5.1}$ &          \citet{Narita2009} &    $61.3^{+6.0}_{-6.8}$ &              This work \\
  HD 189733 b &         $-0.4^{+0.2}_{-0.2}$ &           \citet{Cegla2016} &    $7.0^{+12.0}_{-4.0}$ &      \citet{Cegla2016} \\
  HD 209458 b &          $1.6^{+0.1}_{-0.1}$ & \citet{CasasayasBarris2021} &    $37.3^{+8.8}_{-5.4}$ &              This work \\
    HD 3167 b &         $-6.6^{+6.6}_{-7.9}$ &        \citet{Bourrier2021} &    $29.5^{+7.2}_{-9.4}$ &   \citet{Bourrier2021} \\
    HD 3167 c &       $-108.9^{+5.4}_{-5.5}$ &        \citet{Bourrier2021} &   $107.7^{+5.1}_{-4.9}$ &   \citet{Bourrier2021} \\
  HD 332231 b &         $-2.0^{+6.0}_{-6.0}$ &       \citet{Knudstrup2022} &                     --- &                    --- \\
\hline
\end{tabular}
\begin{tablenotes}
\textit{Note:} as for the 3D spin--orbit angles, we include the values coming for the literature, plus the ones we derive in our study by fitting for $v_{\rm eq} \sin i_\star$ (as described in Sect.~\ref{sec:estim_psi_procedure}). The latter 3D obliquities are referenced as ``This work'' and can be readily used in future studies. We do not include the 3D obliquities we derive based on an isotropic stellar inclination distribution (Sect.~\ref{sec:estim_psi_procedure}) because of the underlying assumptions. We recommend using the latter approach only in ensemble studies and not for individual systems.
\end{tablenotes}
\label{tab:sample}
\end{table*}

\begin{table*}
\caption{Table \ref{tab:sample} continued.}
\centering
\tiny
\begin{tabular}{ccccc}
\hline \hline
  Planet name &           \multicolumn{2}{c}{Projected spin--orbit angle}  &         \multicolumn{2}{c}{3D spin--orbit angle} \\
              &              $\lambda$ (deg) &                      Source &            $\psi$ (deg) &                 Source \\
\hline
   HD 63433 b &        $8.0^{+33.0}_{-45.0}$ &            \citet{Mann2020} &  $33.1^{+17.0}_{-26.2}$ &              This work \\
   HD 63433 c &       $-1.0^{+35.0}_{-32.0}$ &             \citet{Dai2020} &  $26.4^{+13.0}_{-20.3}$ &              This work \\
   HD 80606 b &         $42.0^{+8.0}_{-8.0}$ &         \citet{Hebrard2010} &                     --- &                    --- \\
 HD 85628 A b &       $-115.1^{+2.7}_{-3.6}$ &          \citet{Dorval2020} &                     --- &                    --- \\
   HD 89345 b &       $74.2^{+33.6}_{-32.5}$ &        \citet{Bourrier2023} &  $80.1^{+22.3}_{-23.1}$ &   \citet{Bourrier2023} \\
  HIP 67522 b &          $5.8^{+2.5}_{-3.7}$ &       \citet{Heitzmann2021} &   $20.2^{+10.3}_{-8.7}$ &  \citet{Heitzmann2021} \\
     K2-105 b &      $-81.0^{+50.0}_{-47.0}$ &        \citet{Bourrier2023} &                     --- &                    --- \\
     K2-140 b &          $0.5^{+9.7}_{-9.7}$ &            \citet{Rice2021} &  $24.0^{+10.9}_{-18.5}$ &              This work \\
     K2-232 b &        $-11.1^{+6.6}_{-6.6}$ &            \citet{Wang2021} &                     --- &                    --- \\
      K2-25 b &        $3.0^{+16.0}_{-16.0}$ &      \citet{Stefansson2020} &   $17.0^{+11.0}_{-8.0}$ & \citet{Stefansson2020} \\
     K2-267 b &         $-1.5^{+0.8}_{-0.8}$ &              \citet{Yu2018} &     $3.4^{+3.5}_{-1.6}$ &         \citet{Yu2018} \\
      K2-29 b &          $1.5^{+8.7}_{-8.7}$ &        \citet{Santerne2016} &  $21.6^{+10.3}_{-13.3}$ &              This work \\
     K2-290 b &      $173.0^{+45.0}_{-53.0}$ &          \citet{Hjorth2021} &                     --- &                    --- \\
     K2-290 c &        $153.0^{+8.0}_{-8.0}$ &          \citet{Hjorth2021} &   $124.0^{+6.0}_{-6.0}$ &     \citet{Hjorth2021} \\
      K2-34 b &        $-1.0^{+10.0}_{-9.0}$ &          \citet{Hirano2016} &                     --- &                    --- \\
     KELT-1 b &        $2.0^{+16.0}_{-16.0}$ &          \citet{Siverd2012} &                     --- &                    --- \\
    KELT-11 b &        $-77.9^{+2.4}_{-2.3}$ &         \citet{Mounzer2022} &                     --- &                    --- \\
    KELT-17 b &       $-115.9^{+4.1}_{-4.1}$ &           \citet{Zhou2016b} &   $116.0^{+4.0}_{-4.0}$ &      \citet{Zhou2016b} \\
    KELT-19 b &       $-179.7^{+3.7}_{-3.8}$ &          \citet{Siverd2018} &                     --- &                    --- \\
    KELT-20 b &          $3.4^{+2.1}_{-2.1}$ &            \citet{Lund2017} &  $35.6^{+34.2}_{-34.0}$ &       \citet{Lund2017} \\
    KELT-21 b &         $-5.6^{+1.7}_{-1.9}$ &         \citet{Johnson2017} &                     --- &                    --- \\
    KELT-24 b &          $2.6^{+5.1}_{-3.6}$ &       \citet{Rodriguez2019} &                     --- &                    --- \\
    KELT-25 b &         $23.4^{+3.2}_{-2.3}$ &        \citet{Martinez2020} &                     --- &                    --- \\
   KELT-4 A b &      $14.0^{+100.0}_{-64.0}$ &         \citet{Eastman2016} &                     --- &                    --- \\
     KELT-6 b &      $-36.0^{+11.0}_{-11.0}$ &        \citet{Damasso2015a} &                     --- &                    --- \\
     KELT-7 b &          $2.7^{+0.6}_{-0.6}$ &            \citet{Zhou2016} &                     --- &                    --- \\
     KELT-9 b &        $-84.8^{+0.3}_{-0.3}$ &         \citet{Stephan2022} &    $87.5^{+0.2}_{-0.2}$ &    \citet{Stephan2022} \\
    KOI-142 b &          $6.4^{+0.1}_{-0.1}$ &        \citet{Nesvorny2013} &                     --- &                    --- \\
    KOI-142 c &       $-107.1^{+0.9}_{-0.6}$ &        \citet{Nesvorny2013} &                     --- &                    --- \\
    KOI-368 b &         $10.0^{+2.0}_{-2.0}$ &          \citet{Ahlers2014} &                     --- &                    --- \\
     KOI-94 d &      $-11.0^{+11.0}_{-11.0}$ &        \citet{Albrecht2013} &                     --- &                    --- \\
Kepler-1115 b &        $1.0^{+13.0}_{-13.0}$ &          \citet{Barnes2015} &  $-4.0^{+60.0}_{-60.0}$ &     \citet{Barnes2015} \\
  Kepler-13 b &         $59.2^{+0.1}_{-0.1}$ &         \citet{Howarth2017} &    $60.2^{+0.1}_{-0.1}$ &    \citet{Howarth2017} \\
  Kepler-17 b &        $0.0^{+15.0}_{-15.0}$ &          \citet{Desert2011} &   $0.0^{+15.0}_{-15.0}$ &     \citet{Desert2011} \\
  Kepler-25 c &         $-0.9^{+7.7}_{-6.4}$ &        \citet{Bourrier2023} &    $24.1^{+9.2}_{-9.3}$ &   \citet{Bourrier2023} \\
  Kepler-30 b &        $4.0^{+10.0}_{-10.0}$ &    \citet{SanchisOjeda2012} &                     --- &                    --- \\
 Kepler-420 b &       $74.0^{+32.0}_{-46.0}$ &        \citet{Santerne2014} &                     --- &                    --- \\
 Kepler-448 b &         $-7.1^{+4.2}_{-2.8}$ &         \citet{Johnson2017} &                     --- &                    --- \\
 Kepler-462 b &      $-32.0^{+11.0}_{-11.0}$ &          \citet{Ahlers2015} &    $72.0^{+3.0}_{-3.0}$ &              This work \\
 Kepler-462 c &      $-32.0^{+40.0}_{-40.0}$ &          \citet{Ahlers2015} &    $72.2^{+7.5}_{-9.9}$ &              This work \\
  Kepler-63 b &     $-135.0^{+21.2}_{-26.8}$ &        \citet{Bourrier2023} & $114.6^{+16.6}_{-12.5}$ &   \citet{Bourrier2023} \\
   Kepler-8 b &          $5.0^{+7.0}_{-7.0}$ &        \citet{Albrecht2012} &                     --- &                    --- \\
  Kepler-89 d &       $-6.0^{+13.0}_{-11.0}$ &          \citet{Hirano2012} &                     --- &                    --- \\
   Kepler-9 b &      $-13.0^{+16.0}_{-16.0}$ &            \citet{Wang2018} &                     --- &                    --- \\
  MASCARA-1 b &         $69.2^{+3.1}_{-3.4}$ &          \citet{Hooton2022} &    $72.1^{+2.5}_{-2.4}$ &     \citet{Hooton2022} \\
  MASCARA-4 b &      $244.0^{+15.0}_{-15.0}$ &          \citet{Ahlers2020} &  $104.0^{+7.0}_{-13.0}$ &     \citet{Ahlers2020} \\
     NGTS-2 b &        $-11.3^{+4.8}_{-4.8}$ &        \citet{Anderson2018} &                     --- &                    --- \\
    Qatar-1 b &         $-8.4^{+7.1}_{-7.1}$ &          \citet{Covino2013} &   $18.0^{+7.7}_{-10.8}$ &              This work \\
    Qatar-2 b &          $0.0^{+8.0}_{-8.0}$ &          \citet{Mocnik2017} &                     --- &                    --- \\
   TOI-1268 b &         $40.0^{+7.2}_{-9.9}$ &            \citet{Dong2022} &                     --- &                    --- \\
   TOI-1431 b &     $-155.0^{+20.0}_{-10.0}$ &        \citet{Stangret2021} &                     --- &                    --- \\
   TOI-1518 b &        $240.3^{+0.9}_{-1.0}$ &           \citet{Cabot2021} &                     --- &                    --- \\
   TOI-2025 b &        $9.0^{+33.0}_{-31.0}$ &      \citet{Knudstrup2022b} &                     --- &                    --- \\
   TOI-2109 b &          $1.7^{+1.7}_{-1.7}$ &            \citet{Wong2021} &                     --- &                    --- \\
    TOI-942 b &        $1.0^{+41.0}_{-33.0}$ &           \citet{Wirth2021} &   $2.0^{+27.0}_{-23.0}$ &      \citet{Wirth2021} \\
 TRAPPIST-1 b &       $15.0^{+26.0}_{-30.0}$ &         \citet{Hirano2020b} &                     --- &                    --- \\
 TRAPPIST-1 e &        $9.0^{+45.0}_{-51.0}$ &         \citet{Hirano2020b} &                     --- &                    --- \\
 TRAPPIST-1 f &       $21.0^{+32.0}_{-32.0}$ &         \citet{Hirano2020b} &                     --- &                    --- \\
\hline
\end{tabular}
\end{table*}

\begin{table*}
\caption{Table \ref{tab:sample} continued.}
\centering
\tiny
\begin{tabular}{ccccc}
\hline \hline
     TrES-1 b &       $30.0^{+21.0}_{-21.0}$ &          \citet{Narita2007} &                     --- &                    --- \\
     TrES-2 b &       $-9.0^{+12.0}_{-12.0}$ &           \citet{Winn2008b} &                     --- &                    --- \\
     TrES-4 b &          $6.3^{+4.7}_{-4.7}$ &         \citet{Narita2010b} &                     --- &                    --- \\
  V1298 Tau b &         $4.0^{+7.0}_{-10.0}$ &         \citet{Johnson2022} &     $8.0^{+4.0}_{-7.0}$ &    \citet{Johnson2022} \\
  V1298 Tau c &        $4.9^{+15.0}_{-15.1}$ &       \citet{Feinstein2021} &  $21.3^{+10.5}_{-10.3}$ &              This work \\
     WASP-1 b &      $-59.0^{+99.0}_{-26.0}$ &        \citet{Albrecht2011} &                     --- &                    --- \\
   WASP-100 b &       $79.0^{+19.0}_{-10.0}$ &         \citet{Addison2018} &  $82.0^{+14.9}_{-15.3}$ &              This work \\
   WASP-103 b &        $3.0^{+33.0}_{-33.0}$ &         \citet{Addison2016} &                     --- &                    --- \\
   WASP-107 b &     $-158.0^{+15.2}_{-18.5}$ &        \citet{Bourrier2023} &   $103.5^{+1.7}_{-1.8}$ &   \citet{Bourrier2023} \\
   WASP-109 b &        $99.0^{+10.0}_{-9.0}$ &         \citet{Addison2018} &                     --- &                    --- \\
    WASP-11 b &          $7.0^{+5.0}_{-5.0}$ &         \citet{Mancini2015} &                     --- &                    --- \\
   WASP-111 b &       $-5.0^{+16.0}_{-16.0}$ &        \citet{Anderson2014} &                     --- &                    --- \\
   WASP-117 b &        $-46.9^{+5.5}_{-4.8}$ &          \citet{Carone2021} &    $69.6^{+4.7}_{-4.1}$ &     \citet{Carone2021} \\
    WASP-12 b &       $59.0^{+15.0}_{-20.0}$ &        \citet{Albrecht2012} &                     --- &                    --- \\
   WASP-121 b &         $87.2^{+0.4}_{-0.5}$ &        \citet{Bourrier2020} &    $88.1^{+0.2}_{-0.2}$ &   \citet{Bourrier2020} \\
   WASP-127 b &       $-128.4^{+5.6}_{-5.5}$ &          \citet{Allart2020} &                     --- &                    --- \\
    WASP-13 b &        $8.0^{+13.0}_{-12.0}$ &       \citet{Brothwell2014} &                     --- &                    --- \\
   WASP-134 b &        $-43.7^{+9.9}_{-9.9}$ &       \citet{Anderson2018b} &                     --- &                    --- \\
    WASP-14 b &        $-33.1^{+7.4}_{-7.4}$ &         \citet{Johnson2009} &                     --- &                    --- \\
   WASP-148 b &         $-8.2^{+8.7}_{-9.7}$ &            \citet{Wang2022} &  $29.8^{+13.2}_{-23.4}$ &              This work \\
    WASP-15 b &       $-139.6^{+5.2}_{-4.3}$ &          \citet{Triaud2010} &                     --- &                    --- \\
   WASP-156 b &      $105.7^{+14.0}_{-14.4}$ &        \citet{Bourrier2023} &                     --- &                    --- \\
    WASP-16 b &       $-4.2^{+11.0}_{-13.9}$ &          \citet{Brown2012b} &                     --- &                    --- \\
   WASP-166 b &         $-0.7^{+1.6}_{-1.6}$ &        \citet{Bourrier2023} &   $14.2^{+6.0}_{-12.8}$ &              This work \\
   WASP-167 b &       $-165.0^{+5.0}_{-5.0}$ &          \citet{Temple2017} & $145.6^{+17.6}_{-11.6}$ &              This work \\
    WASP-17 b &       $-148.5^{+5.1}_{-4.2}$ &          \citet{Triaud2010} &                     --- &                    --- \\
   WASP-174 b &         $31.0^{+1.0}_{-1.0}$ &          \citet{Temple2018} &                     --- &                    --- \\
   WASP-178 b &         $91.3^{+6.5}_{-6.3}$ &        \citet{Martinez2020} &                     --- &                    --- \\
    WASP-18 b &          $4.0^{+5.0}_{-5.0}$ &          \citet{Triaud2010} &                     --- &                    --- \\
   WASP-180 b &       $-162.0^{+5.0}_{-5.0}$ &          \citet{Temple2019} &   $148.9^{+9.6}_{-7.8}$ &              This work \\
   WASP-189 b &         $91.7^{+1.2}_{-1.2}$ &          \citet{Deline2022} &    $89.6^{+1.2}_{-1.2}$ &     \citet{Deline2022} \\
    WASP-19 b &         $-1.9^{+1.1}_{-1.1}$ &       \citet{Sedaghati2021} &   $14.2^{+7.8}_{-12.9}$ &              This work \\
   WASP-190 b &         $21.0^{+6.0}_{-6.0}$ &         \citet{Temple2019b} &                     --- &                    --- \\
     WASP-2 b &     $-153.0^{+11.0}_{-15.0}$ &          \citet{Triaud2010} &                     --- &                    --- \\
    WASP-20 b &         $12.7^{+4.2}_{-4.2}$ &       \citet{Anderson2015b} &                     --- &                    --- \\
    WASP-21 b &        $8.0^{+26.0}_{-27.0}$ &            \citet{Chen2020} &                     --- &                    --- \\
    WASP-22 b &       $22.0^{+16.0}_{-16.0}$ &        \citet{Anderson2011} &  $30.5^{+13.6}_{-14.3}$ &              This work \\
    WASP-24 b &         $-4.7^{+4.0}_{-4.0}$ &         \citet{Simpson2011} &                     --- &                    --- \\
    WASP-25 b &         $14.6^{+6.7}_{-6.7}$ &          \citet{Brown2012b} &                     --- &                    --- \\
    WASP-26 b &      $-34.0^{+36.0}_{-26.0}$ &        \citet{Albrecht2012} &                     --- &                    --- \\
    WASP-28 b &        $8.0^{+18.0}_{-18.0}$ &       \citet{Anderson2015b} &                     --- &                    --- \\
     WASP-3 b &          $5.0^{+6.0}_{-5.0}$ &          \citet{Miller2010} &                     --- &                    --- \\
    WASP-31 b &          $2.8^{+3.1}_{-3.1}$ &          \citet{Brown2012b} &                     --- &                    --- \\
    WASP-32 b &         $10.5^{+6.4}_{-6.5}$ &           \citet{Brown2012} &   $2.0^{+16.0}_{-16.0}$ &      \citet{Brown2012} \\
    WASP-33 b &       $-111.6^{+0.3}_{-0.3}$ &        \citet{Watanabe2022} &   $108.2^{+0.9}_{-1.0}$ &   \citet{Watanabe2022} \\
    WASP-38 b &          $7.5^{+4.7}_{-6.1}$ &           \citet{Brown2012} &                     --- &                    --- \\
    WASP-39 b &        $0.0^{+11.0}_{-11.0}$ &         \citet{Mancini2018} &                     --- &                    --- \\
     WASP-4 b &        $4.0^{+43.0}_{-34.0}$ &          \citet{Triaud2010} &  $39.3^{+16.8}_{-24.2}$ &              This work \\
    WASP-41 b &        $6.0^{+11.0}_{-11.0}$ &      \citet{Southworth2016} &    $16.7^{+7.9}_{-9.3}$ &              This work \\
    WASP-43 b &          $3.5^{+6.8}_{-6.8}$ &        \citet{Esposito2017} &                     --- &                    --- \\
    WASP-47 b &        $0.0^{+24.0}_{-24.0}$ &    \citet{SanchisOjeda2015} &  $29.2^{+11.1}_{-13.3}$ &   \citet{Bourrier2023} \\
    WASP-49 b &       $54.0^{+79.0}_{-58.0}$ &      \citet{Wyttenbach2017} &                     --- &                    --- \\
     WASP-5 b &        $12.1^{+10.0}_{-8.0}$ &          \citet{Triaud2010} &  $26.6^{+12.5}_{-14.9}$ &              This work \\
    WASP-52 b &          $1.1^{+1.1}_{-1.1}$ &           \citet{Chen2020b} &                     --- &                    --- \\
    WASP-53 b &       $-1.0^{+12.0}_{-12.0}$ &          \citet{Triaud2017} &                     --- &                    --- \\
     WASP-6 b &          $7.2^{+3.7}_{-3.7}$ &    \citet{TregloanReed2015} &  $29.6^{+13.7}_{-22.7}$ &              This work \\
    WASP-60 b &     $-129.0^{+17.0}_{-17.0}$ &         \citet{Mancini2018} &                     --- &                    --- \\
    WASP-61 b &        $4.0^{+17.1}_{-18.4}$ &           \citet{Brown2017} &                     --- &                    --- \\
    WASP-62 b &         $19.4^{+5.1}_{-4.9}$ &           \citet{Brown2017} &                     --- &                    --- \\
\hline
\end{tabular}
\end{table*}

\begin{table*}
\caption{Table \ref{tab:sample} continued.}
\centering
\tiny
\begin{tabular}{ccccc}
\hline \hline
    WASP-66 b &       $-4.0^{+22.0}_{-22.0}$ &         \citet{Addison2016} &                     --- &                    --- \\
    WASP-69 b &          $0.4^{+2.0}_{-1.9}$ & \citet{CasasayasBarris2017} &   $15.6^{+7.3}_{-14.3}$ &              This work \\
     WASP-7 b &         $86.0^{+6.0}_{-6.0}$ &       \citet{Albrecht2012b} &                     --- &                    --- \\
    WASP-71 b &         $-1.9^{+7.1}_{-7.5}$ &           \citet{Brown2017} &                     --- &                    --- \\
    WASP-72 b &       $-7.0^{+11.0}_{-12.0}$ &         \citet{Addison2018} &                     --- &                    --- \\
    WASP-74 b &          $0.8^{+1.0}_{-1.0}$ &           \citet{Luque2020} &                     --- &                    --- \\
    WASP-76 b &         $61.3^{+7.6}_{-5.1}$ &      \citet{Ehrenreich2020} &                     --- &                    --- \\
    WASP-78 b &         $-6.4^{+5.9}_{-5.9}$ &           \citet{Brown2017} &                     --- &                    --- \\
    WASP-79 b &        $-99.1^{+4.1}_{-3.9}$ &         \citet{Johnson2017} &                     --- &                    --- \\
     WASP-8 b &       $-143.0^{+1.6}_{-1.5}$ &        \citet{Bourrier2017} &                     --- &                    --- \\
    WASP-80 b &      $-14.0^{+14.0}_{-14.0}$ &          \citet{Triaud2015} &                     --- &                    --- \\
    WASP-84 b &         $-0.3^{+1.7}_{-1.7}$ &        \citet{Anderson2015} &    $17.3^{+7.7}_{-7.7}$ &   \citet{Anderson2015} \\
    WASP-87 b &       $-8.0^{+11.0}_{-11.0}$ &         \citet{Addison2016} &                     --- &                    --- \\
    WASP-94 b &      $151.0^{+16.0}_{-23.0}$ &   \citet{NeveuVanMalle2014} &                     --- &                    --- \\
       XO-2 b &        $7.0^{+11.0}_{-11.0}$ &        \citet{Damasso2015b} &  $27.0^{+12.0}_{-27.0}$ &   \citet{Damasso2015b} \\
       XO-3 b &         $37.3^{+3.0}_{-3.0}$ &          \citet{Hirano2011} &                     --- &                    --- \\
       XO-4 b &        $-46.7^{+8.1}_{-6.1}$ &          \citet{Narita2010} &                     --- &                    --- \\
       XO-6 b &        $-20.7^{+2.3}_{-2.3}$ &         \citet{Crouzet2017} &                     --- &                    --- \\
     pi Men c &        $-24.0^{+4.1}_{-4.1}$ &         \citet{Kunovac2021} &    $26.9^{+5.8}_{-4.7}$ &    \citet{Kunovac2021} \\
\hline
\end{tabular}
\end{table*}

\end{appendix}


\end{document}